\documentclass{aa}  

\usepackage{graphicx}
\usepackage{txfonts}
\usepackage{hyperref}
\usepackage{natbib}
\bibpunct{(}{)}{;}{a}{}{,} 

\begin{document}

\title{A 3D view of the Taurus star-forming region by {\it Gaia} and {\it Herschel}: }
\subtitle{multiple populations related to the filamentary molecular cloud}
   \author{V.~Roccatagliata\inst{1,2,3}, E. Franciosini\inst{2},  G.~G.~Sacco\inst{2}, S.~Randich\inst{2} \and A. Sicilia-Aguilar\inst{4}
          }

\institute{Dipartimento di Fisica ``Enrico Fermi'', Universita' di Pisa, Largo Pontecorvo 3, 56127 Pisa, Italy  \\
\email{veronica.roccatagliata@unipi.it} 
\and INAF-Osservatorio Astrofisico di Arcetri, Largo E. Fermi 5, 50125 Firenze, Italy
\and INFN, Sezione di Pisa, Largo Bruno Pontecorvo 3, 56127 Pisa,
Italy
\and SUPA, School of Science and Engineering, University of Dundee, Nethergate, DD1 4HN, Dundee, UK}
   \date{Received 29 July 2019; Accepted 18 April 2020}

  \abstract
   {Taurus represents an ideal region to study the  three-dimensional distribution of the young stellar population and relate it to the associated molecular cloud.
  }
   {The second {\it Gaia} data release (DR2) enables us to investigate the Taurus complex in three dimensions, starting from a previously defined robust membership. The molecular cloud structured in filaments can be traced in emission using the public far-infrared maps from {\it Herschel}. 
    }
   {From a compiled catalog of spectroscopically confirmed members, we analyze the 283 sources with reliable parallax and proper motions in the {\it Gaia} DR2 archive. 
   We fit the distribution of parallaxes and proper motions with multiple populations described by multivariate Gaussians. 
  We compute the cartesian Galactic coordinates (X,Y,Z) and, for the populations 
  associated with the main cloud,
   also the galactic space velocity (U,V,W).
   We discuss the spatial distribution of the populations in relation to the structure of the filamentary molecular cloud traced by {\it Herschel}.
   }
   {
   We discover the presence of six populations which are all well defined in parallax and proper motions, with the only exception being Taurus D. The derived distances range  between $\sim$130 and $\sim$160 pc. We do not find a unique relation between stellar population and the associated molecular cloud: while the stellar population seems to be on the cloud surface,  both lying at similar distances, this is not the case when the molecular cloud is structured in filaments. Taurus B is  probably moving in the direction of Taurus A, while Taurus E appears to be moving towards them. 
   }
 { The Taurus region is the result of a complex star formation history which most probably occurred in clumpy and filamentary structures
that are evolving independently.}

   \keywords{Open clusters and associations: individual: \object{Taurus} - Stars: pre-main sequence -  Parallaxes - Proper motions
               }

   \titlerunning{A 3D view of the Taurus star-forming region by {\it Gaia} and {\it Herschel}} 
   \authorrunning{V. Roccatagliata et al.}

   \maketitle
   %

\section{Introduction}

 During the early phases of the star and cluster formation process, filamentary structures have not only  been theoretically predicted  \citep[e.g., ][]{Daleetal2012a}, but have also been detected in low- and high-mass
star-forming regions, and in regions where no star formation is currently active \citep[e.g., ][]{Andreetal2010}. 
In particular, \citet{Myers2009} highlights that near young stellar groups or clusters are associated with a ``hub-filament structure''. 
Filaments traced commonly by extinction maps  from infrared observations can now be traced in great detail in emission 
in the far-infrared thanks to {\it Herschel}. 
Within 300 pc we find examples of a main prominent filament in  Chamaeleon I or Corona Australis \citep[e.g., ][]{Schmalzletal2010, Palmeirimetal2013, Sicilia-Aguilaretal2011, Sicilia-Aguilaretal2013};  Lupus I is instead located along a filament at the converging location of two bubbles \citep[e.g., ][]{Gaczkowskietal2015, Gaczkowskietal2017, Krauseetal2018}, while Serpens Main is at the converging point of filaments \citep[e.g., ][]{Roccatagliataetal2015}. \\
A step further in the study of filamentary  molecular clouds is to relate their physical properties to young stellar populations. This can be done 
by combining the astrometric {\it Gaia} data with the {\it Herschel} maps of the molecular cloud in emission. The second release of {\it Gaia} is revolutionizing the study of star formation and young clusters under different aspects. 
A 3D density map for early-type and pre-main sequence sources was recently constructed  by,  for example, \citet{Zarietal2018}, finding that 
younger stars within 500 pc to the Sun are mostly distributed  in dense and compact clumps, while older sources are instead widely distributed. Several 
studies on large-scale star-forming regions are combining {\it Gaia} data with photometric surveys  \citep[e.g., Vela OB2, ][]{Armstrongetal2018, Cantat-Gaudinetal2018a}.  New members have been found in nearby 
associations  \citep[e.g.,][]{Gagneetal2018a}, and even new nearby associations have been discovered 
\citep[e.g.,][]{Gagneetal2018b}. Recently, multiple populations were also detected in young clusters \citep{Roccatagliataetal2018, Franciosinietal2018}.\\
\noindent 
One of the most studied young star forming regions, where both a prominent molecular cloud is structured in 
a main filament and a young population is present, is the Taurus complex. 
\noindent
The discussion of star formation history in space and time started more than 15 years ago. 
\citet{Palla&Stahler2002} found a younger  population inside the filaments and a more dispersed and older population outside them, and concluded that an age spread was present in the region. 
In the same year, overlaying the young stellar content to the $^{12}$CO maps, \citet{Hartmann2002} found that most of the sources in {\it ``Taurus Main''} follow three near parallel elongated bands, while only a few of them are localized in widely distributed groups. 
This led him to organize the sources into different groups according to their spatial distribution, without finding a significant age spread in the K7-M1 spectral type range. 
Continuum and line maps of the region with increasing spatial resolution have been obtained for Taurus in recent years. 
A large-scale survey of Taurus in $^{12}$CO and $^{13}$CO was presented by \citet{Goldsmithetal2008}, showing a very complex and highly 
structured cloud morphology including filaments, cavities, and rings. Analyzing the  $^{13}$CO integrated intensities in 
three different velocity intervals, these latter authors found the highest velocity (7-9 km/s) corresponding to L1506 and L1498 and part of L1495, while the other clouds (B213, L1521, B18, L1536) have mainly intermediate velocities (5-7 km/s) and only a few 
points at the borders of those clouds are found at even lower velocities (3-5 km/s). A different orientation of the magnetic field was also found: the
long filament of L1506 is parallel to the field, while the magnetic field is perpendicular to the extension of the long axis of the B216 and B217 filaments 
\citep{Goodmanetal1992}. \\
\noindent
The main filament in the B213-L1495 region has been extensively studied by \citet{Hacaretal2013} in high-density tracers such as C$^{18}$O,  N$_2$H$^+$,
 and SO, and by \citet{Hacaretal2016}  in the three main isotopologs of $^{12}$CO, $^{13}$CO, and C$^{18}$O. 
 These latter authors suggested that the velocity broadening of the lines could be caused by the effects of different structures in 3D, which are clumpy or tangly on the line
 of sight projection of the velocity component.
 \begin{figure}
\centering%
\resizebox{\hsize}{!}{\includegraphics[trim=0cm 0cm 0cm 0cm,width=0.3\textwidth]{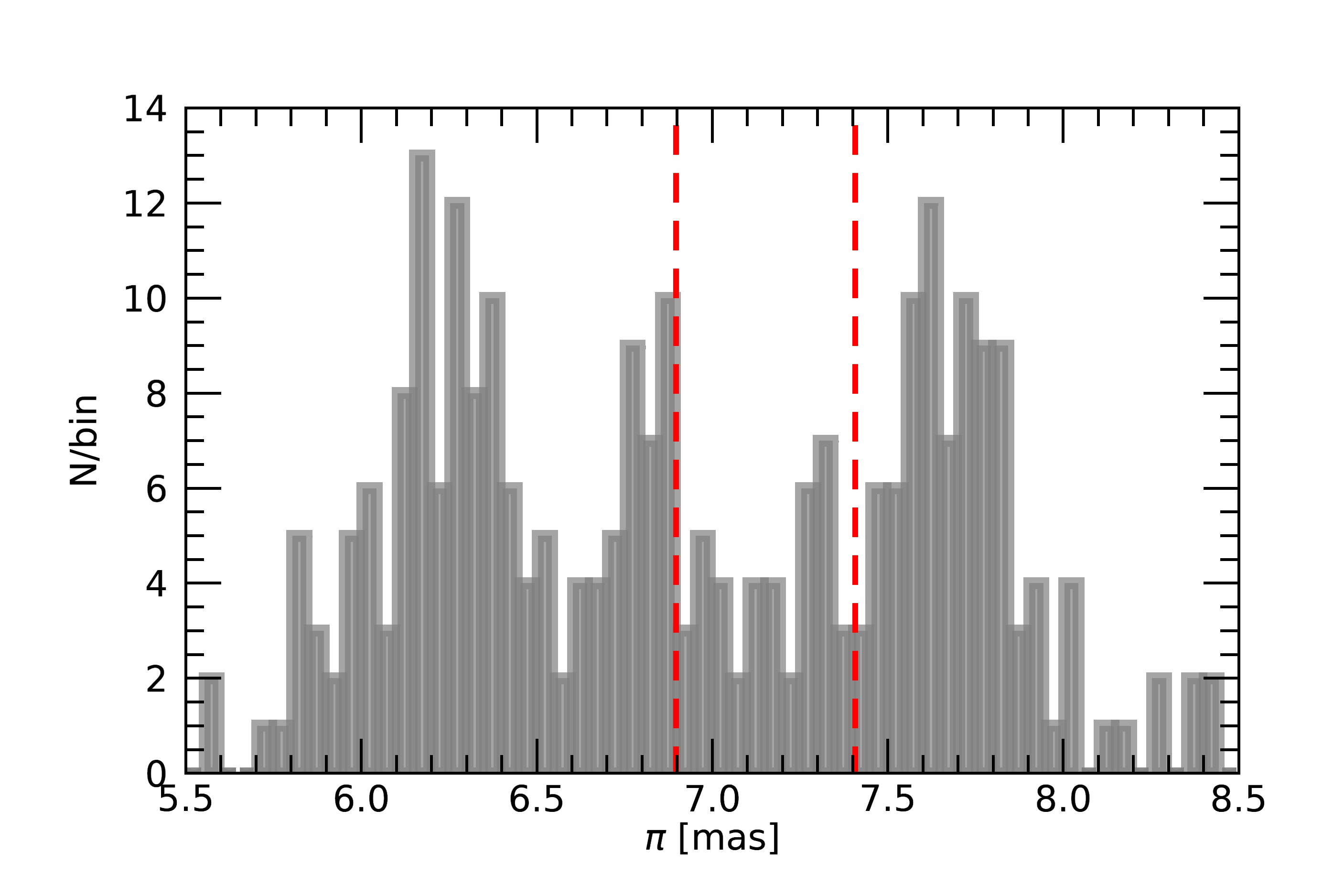}} 
\caption{Histogram of the parallaxes of the Taurus members. The dashed red lines indicate the range in parallax commonly adopted in the literature.}
\label{histplx}
\end{figure}
\noindent
A first analysis of the {\it Gaia} DR2 data of Taurus recently presented by \citet{Luhman2018} and \citet{EsplinLuhman2019} enabled these authors to revise the membership and to constrain the initial mass function of Taurus. Using the {\it Gaia} data, \citet{Luhman2018} 
 found that the population older that 10 Myr is not associated to Taurus. 
\begin{figure}[htb]
\centering%
\resizebox{\hsize}{!}{\includegraphics[trim=2cm 6cm 0cm 0cm,width=0.4\textwidth]{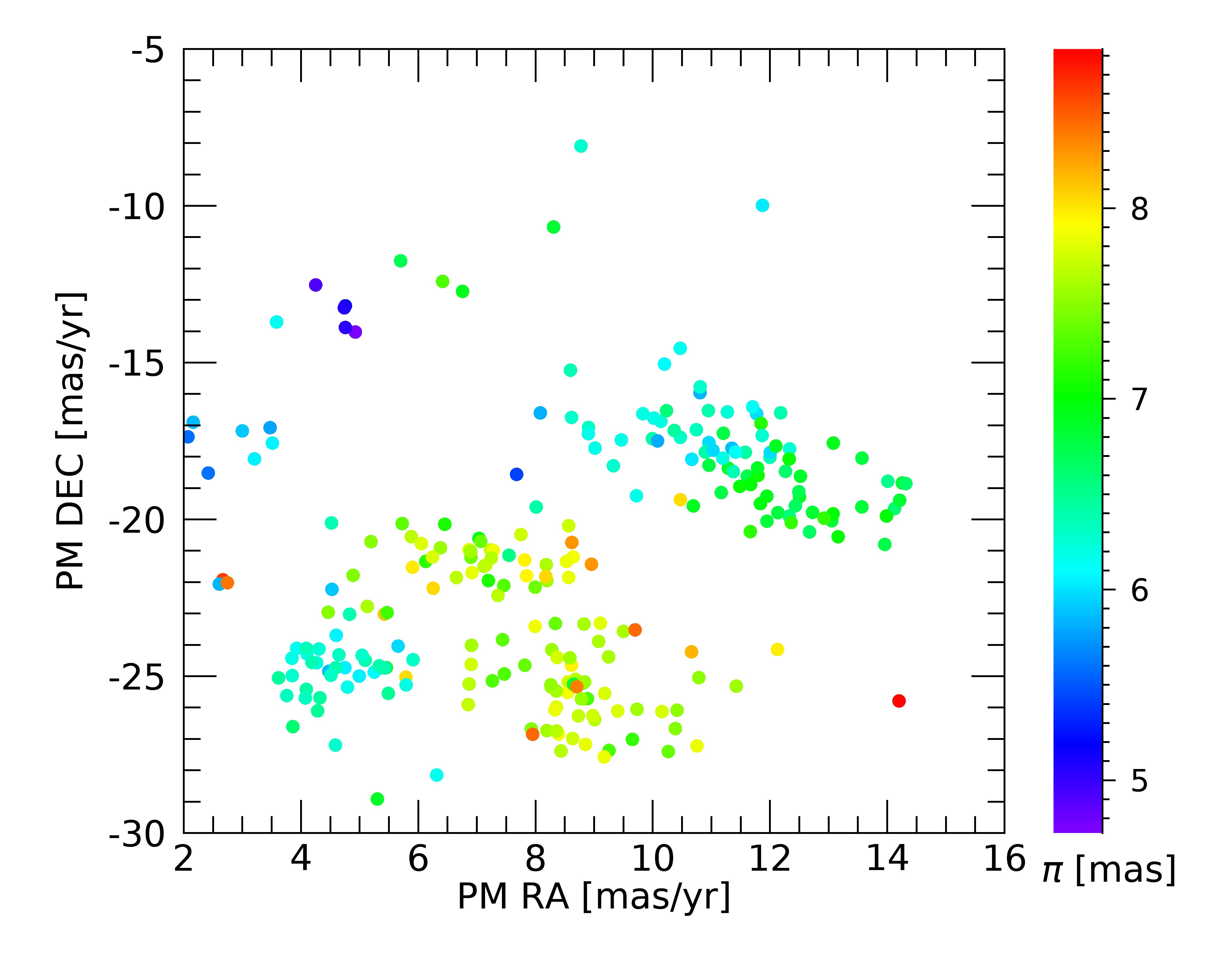}}
\caption{Proper motions in $\alpha$  and $\delta$  color coded by the parallax of the sources.}
\label{pm}
\end{figure}
The goal of the present paper is to relate the star formation history in the Taurus complex to the filamentary structure of the molecular cloud. To this aim, 
the new analysis of the {\it Gaia} DR2 data 
starts from a reliable membership (presented in Sect.~\ref{memb}) and applies a rigorous 
approach which takes into account the covariances \citep[not considered in the studies of][]{Luhman2018, EsplinLuhman2019} between parallax and proper motions (available in the {\it Gaia} Archive). 
The cluster kinematics and  the relation between stellar content and filamentary structures 
are discussed in Sect.~\ref{discussion}, together with the estimate of the distance of the molecular cloud.  Our conclusions are presented in Sect.~\ref{concl}. 
 \section{Cluster membership and {\it Gaia} DR2 data}
\label{memb}
 Of the initial catalog of 518 spectroscopically confirmed members of \citet{EsplinLuhman2019},  443 sources are present in the {\it Gaia} DR2 archive.  
From this sample, we selected the sources with good astrometry following \citet{Lindegren2018tn}, who suggested the use of renormalized unit weight error (RUWE),   which is a more reliable and informative goodness-of-fit statistic than the astrometric excess noise commonly used until now by the community. The RUWE is defined as
\begin{equation}
    RUWE = \frac{\sqrt{\chi^2/(nObs - \nu)}}{f(G,G_{BP}-G_{RP})}
,\end{equation}
where $nObs$ is the number of observations, $\nu$ is the number of parameters solved, and $f$ is a renormalization function. Our  further analysis is 
based on the 283 sources with RUWE of less than 1.4.\\
Figure~\ref{histplx} shows the distribution of the parallaxes of the cluster members. 
The resulting mean parallax is  7.01$\pm$0.06 mas, where the associated error is computed as the standard deviation (0.78 mas) divided by $\sqrt{N}$. 
However, this mean parallax value lies exactly in the middle of a parallax distribution that appears to be multimodal.  
Furthermore, the distribution in proper motions in Fig.~\ref{pm} shows that multiple populations might be present in the region.
A deeper analysis of the {\it Gaia} data is therefore necessary to understand the star formation history of the region.

\section{Analysis}
\label{an1}
 Figures \ref{histplx} and \ref{pm} clearly show that Taurus cannot be described by a single population because both parallaxes and proper motions are not randomly distributed around single mean values. Therefore, 
we model the distribution of the astrometric parameters using a similar approach to that used by \citet{Roccatagliataetal2018} to study the two sub-populations in Chamaeleon I. We assume that the Taurus star forming region is composed of multiple populations, each described by a 3D multivariate Gaussian \citep[as in ][\footnote{Equations 6-10}]{lindegrenetal2000}. The $j$-th population is defined by seven parameters: the mean parallax ($\pi_{j}$), the mean proper motions along right ascension and declination ($\mu_{\alpha,j}$ and $\mu_{\delta,j}$), the intrinsic dispersion of the parallax ($\sigma_{\pi,j}$), the intrinsic dispersion of proper motions along right ascension and declination($\sigma_{\mu_{\alpha},j}$ and $\sigma_{\mu_{\delta},j}$), and a normalization factor that takes into account the fraction of stars belonging to that population ($f_{j}$).  We assume that the mean proper motions and the mean parallaxes of each population are not correlated.
To fit our distribution we use a maximum likelihood technique as in \cite{Jeffriesetal2014} and \cite{Franciosinietal2018},
 adopting the {\sc slsqp} minimization method as implemented in {\tt scipy}. 
The likelihood function of the $i$-th star for the $j$-th population is given by   
 \smallskip
 
\begin{equation}
\label{pdf}
 L_{i,j} = (2\pi)^{-3/2}\,|C_{i,j}|^{-1/2} \times \exp{\begin{bmatrix} -\frac{1}{2}(a_i-a_j)' \,C_{i,j}^{-1}\,(a_j-a_i) \end{bmatrix}}
 ,\end{equation}
\noindent
where $C_{i,j}$ is the covariance matrix, $|C_{i,j}|$ its determinant, and  \\
\noindent
$(a_i-a_j,)'$ is the transpose of the vector 
\begin{equation}
\label{term}
a_i-a_j = \begin{bmatrix} 
\pi_i - \pi_j,\\
\mu_{\alpha,i} - \mu_{\alpha, j}\\
\mu_{\delta,i} - \mu_{\delta, j}
\end{bmatrix}
.\end{equation}
 Following \citet{lindegrenetal2000}, the covariance matrix  $C_{i,j}$ is given by:
\begin{equation}
\small
 C_i\,=\,
 \begin{bmatrix} 
 \sigma_{\pi,i}^2+\sigma_{\pi,j}^2&  \sigma_{\pi,i}\,\sigma_{\mu_{\alpha,i}}\, \rho_i\,(\pi, \mu_{\alpha})& \sigma_{\pi,i}\,\sigma_{\mu_{\delta,i}}\,\rho_i\,(\pi, \mu_{\delta}) \\
 \sigma_{\pi,i}\,\sigma_{\mu_{\alpha,i}}\,\rho_i\,(\pi, \mu_{\alpha})&  \sigma_{\mu_{\alpha,i}}^2+\sigma_{\mu_{\alpha,j}}^2 &  \sigma_{\mu_{\alpha,i}}\,\sigma_{\mu_{\delta,i}}  \cdot \rho_i\,(\mu_{\alpha}, \mu_{\delta})\\
 \sigma_{\pi,i}\,\sigma_{\mu_{\delta,i}}\,\rho_i\,(\pi, \mu_{\delta})&  \sigma_{\mu_{\alpha,i}}\,\sigma_{\mu_{\delta,i}}\,\rho_i\,(\mu_{\alpha}, \mu_{\delta})& \sigma_{\mu_{\delta,i}}^2+\sigma_{\mu_{\delta,j}}^2
\end{bmatrix}
,\end{equation}
%
where $\rho_i\,(\pi, \mu_{\alpha})$, $ \rho_i\,(\pi, \mu_{\delta})$, $\rho_i\,(\mu_{\alpha}, \mu_{\alpha})$ are the  non-diagonal elements of the covariance matrix of each star provided by the {\it Gaia} archive. 

\begin{figure}
\centering%
\includegraphics[trim=0cm 0cm 0cm 0cm,width=0.45\textwidth]{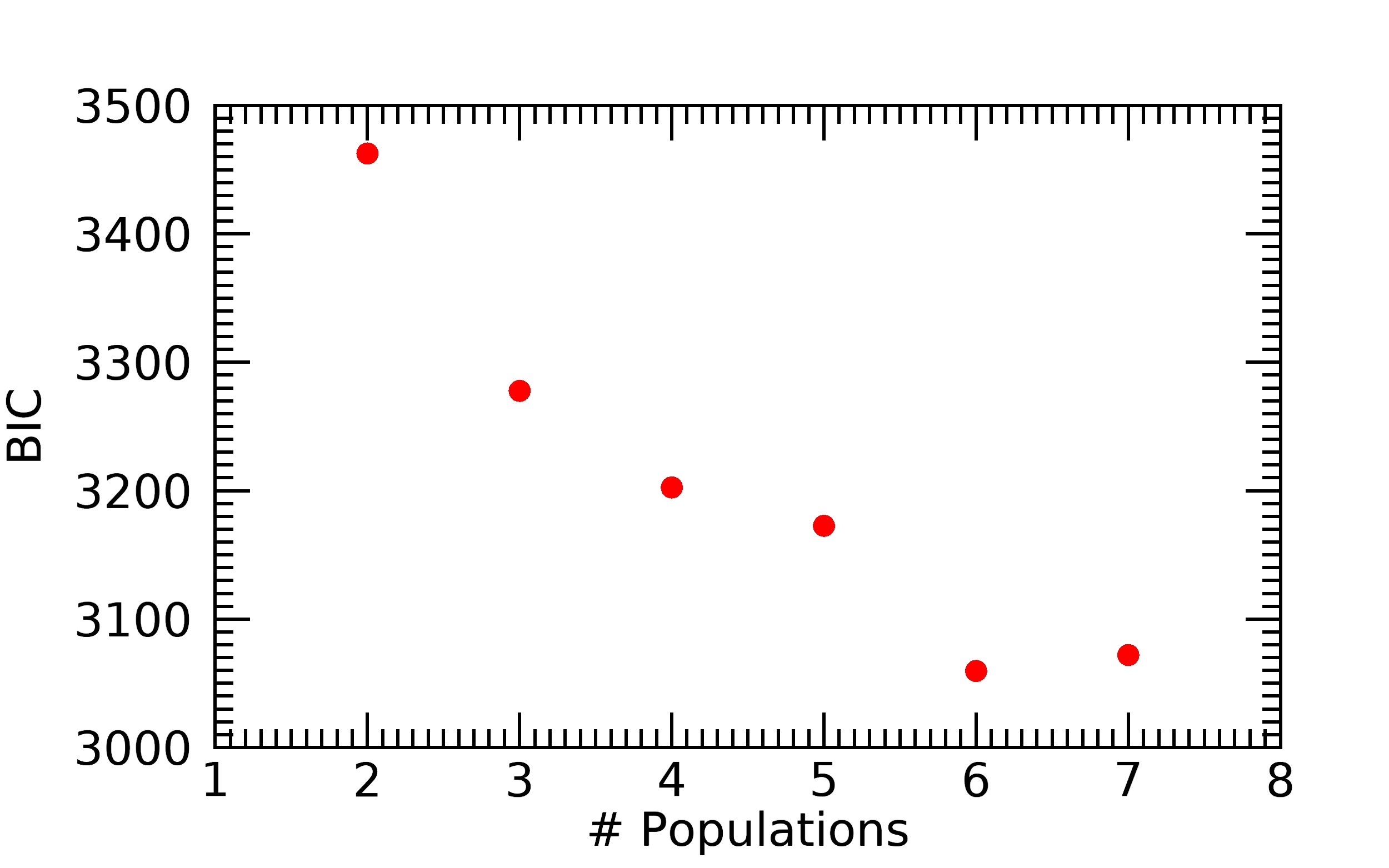} 
    \caption{Values of the BIC calculated for the fits with different numbers $n$ of populations. }
\label{BIC}
\end{figure}

The total likelihood for the $i$-th star is given by:
\begin{equation}
\label{2pop}
 L_i = \sum_{j=1}^{n} f_j\,L_{i,j}
 ,\end{equation}
where $n$ is the number of populations and $\sum_{j=1}^n f_j =1$. 
The probability for each star of belonging to the $j$-th population is then
\begin{equation}
\label{probability}
 \smallskip
P_{i,j}\,=\,f_{j}\,\frac{L_{i,j}}{L_i}  
 .\end{equation}

\begin{figure}
\centering%
\includegraphics[trim=3cm 3.5cm 2cm 3cm,width=0.4\textwidth]{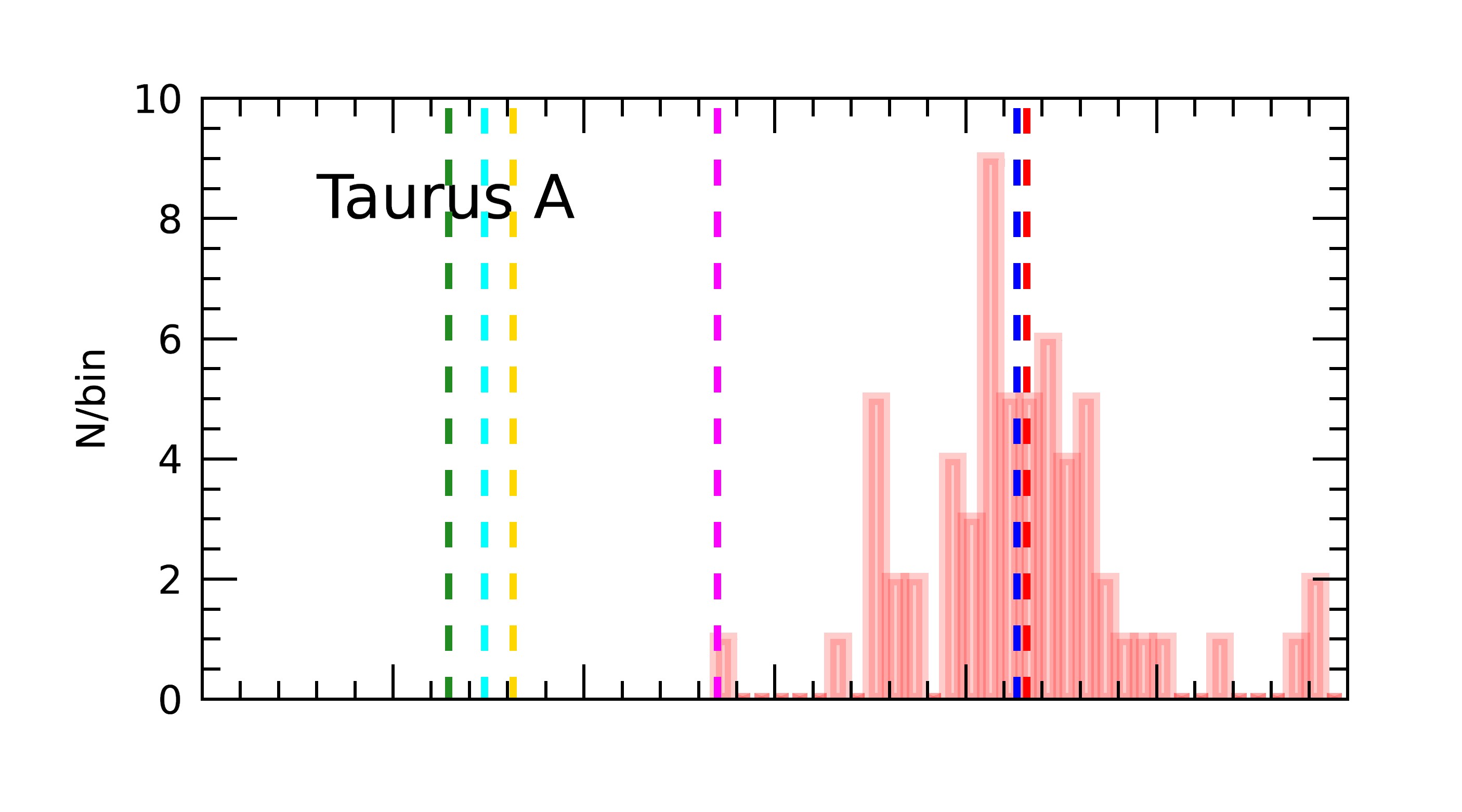}     
\includegraphics[trim=3cm 3.5cm 2cm 3.2cm,width=0.4\textwidth]{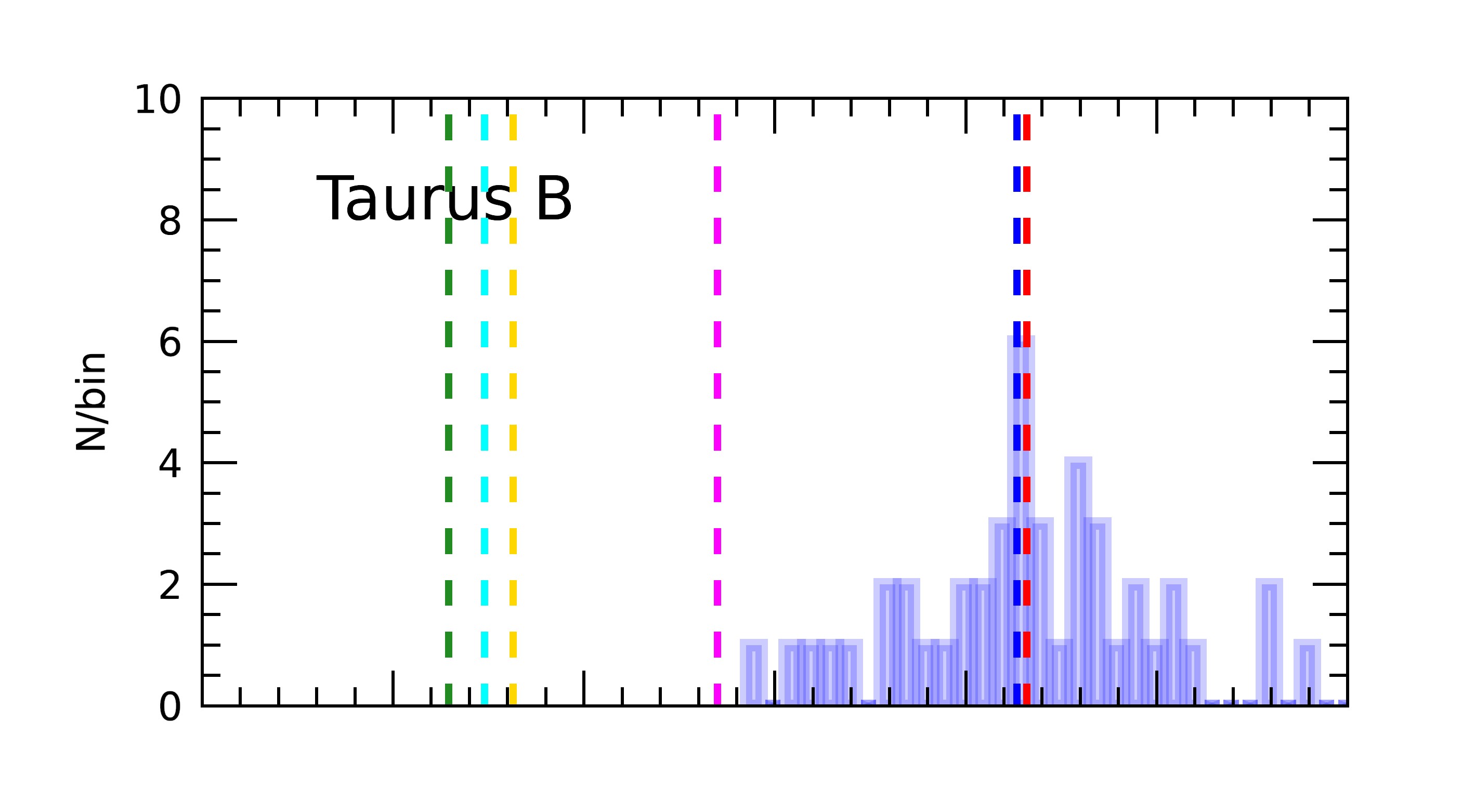}\\
\includegraphics[trim=3cm 3.5cm 2cm 3.2cm,width=0.4\textwidth]{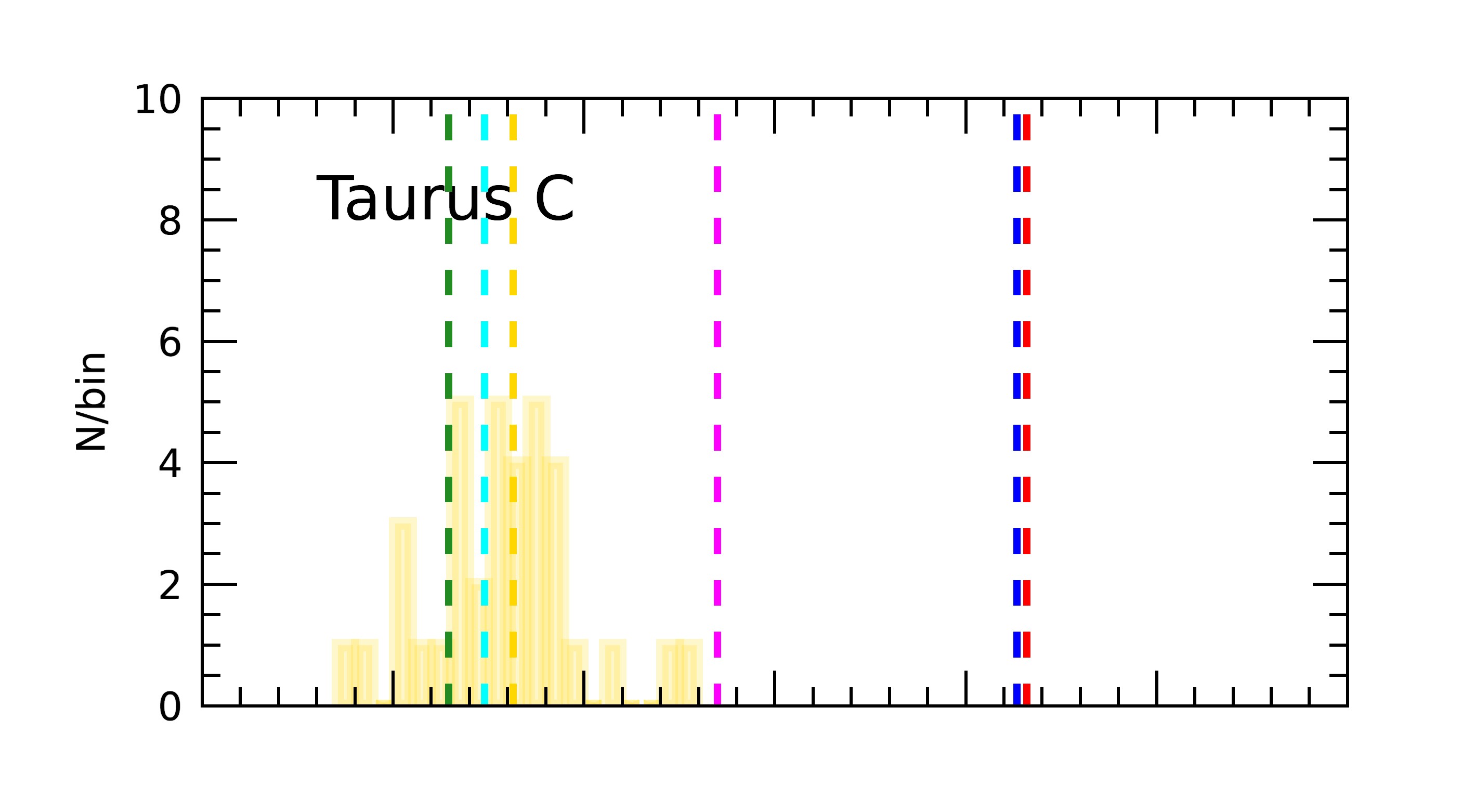}\\
\includegraphics[trim=3cm 3.5cm 2cm 3.2cm,width=0.4\textwidth]{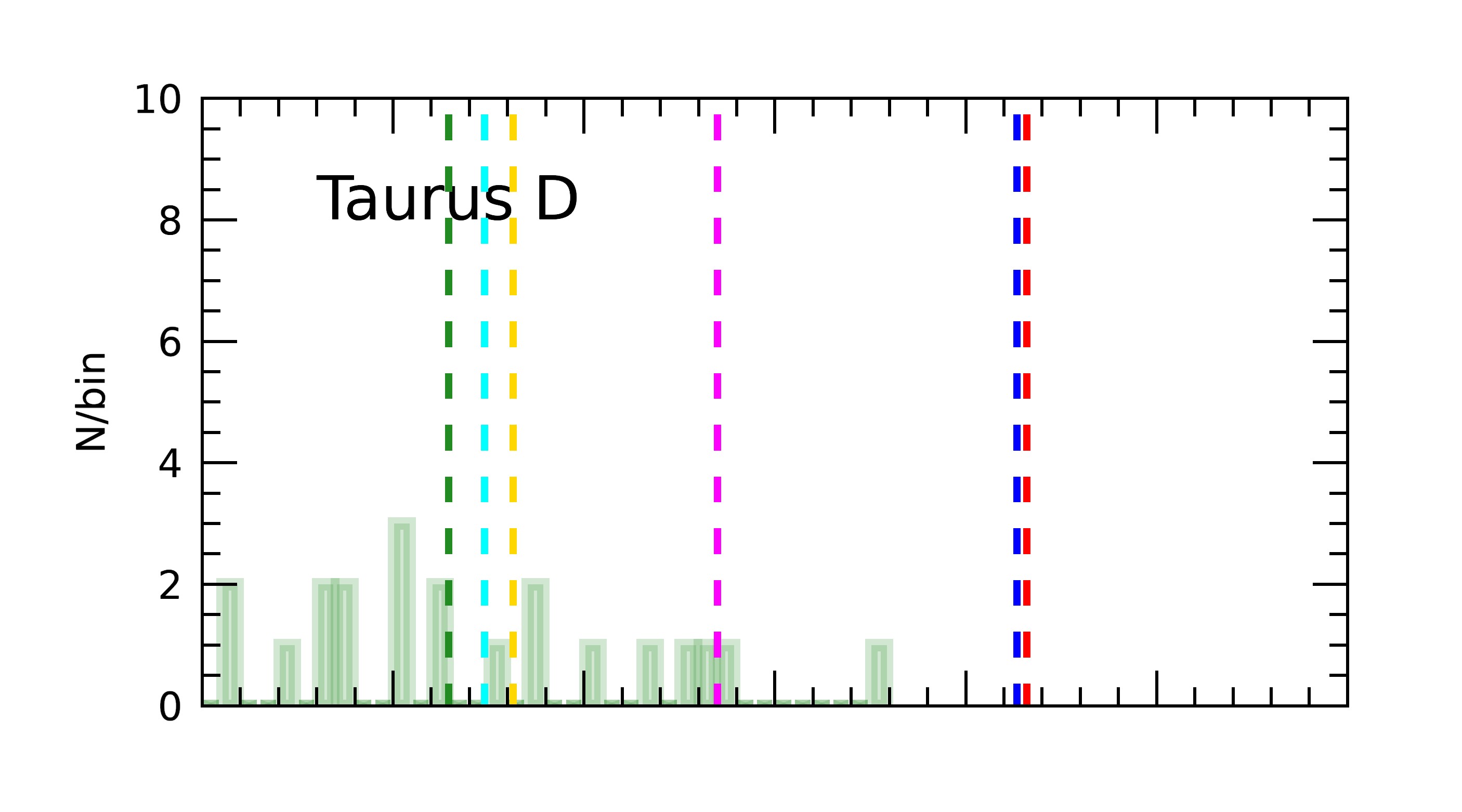}
\includegraphics[trim=3cm 3.5cm 2cm 3.2cm,width=0.4\textwidth]{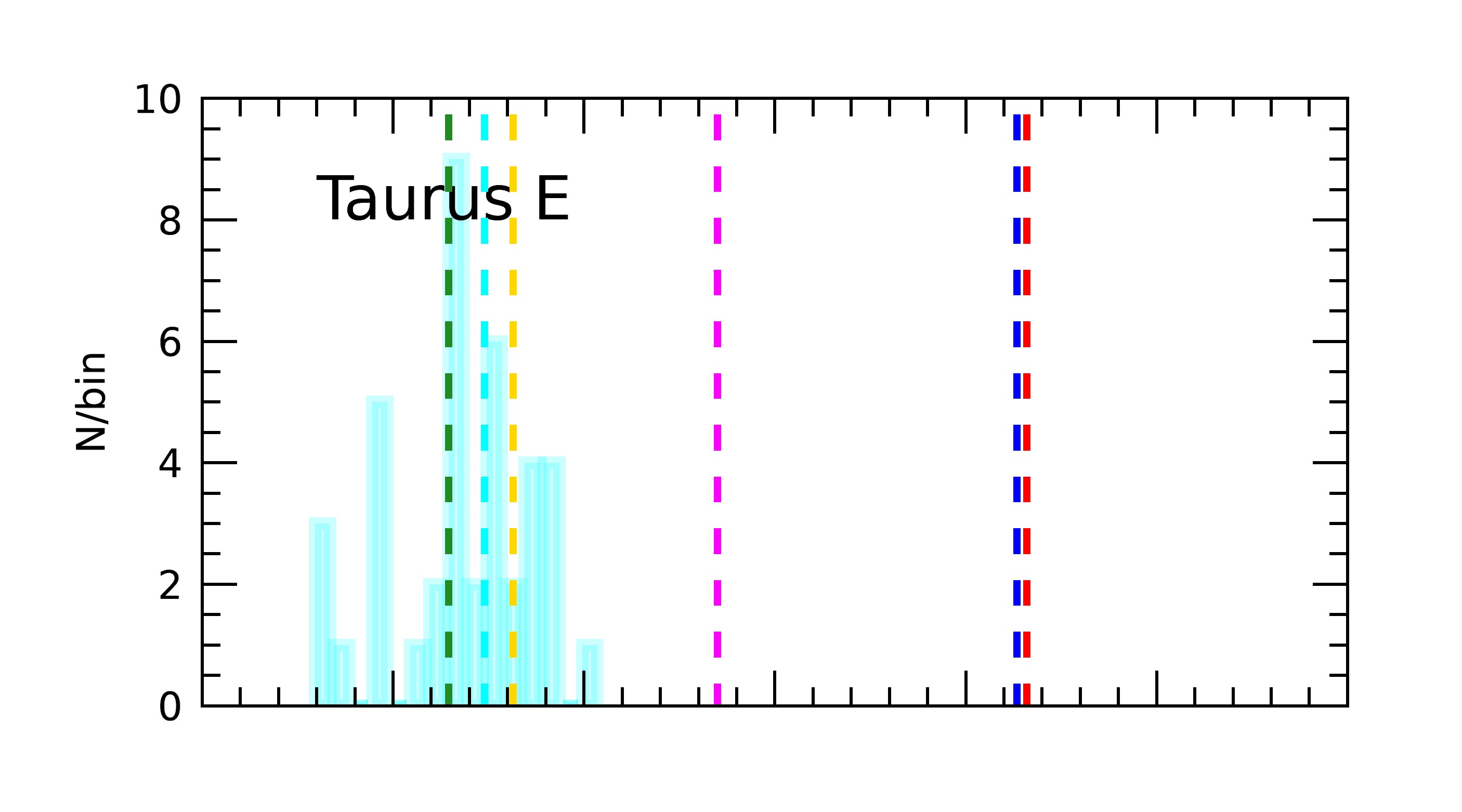}\\
\includegraphics[trim=3cm 3.5cm 2cm 3.2cm,width=0.4\textwidth]{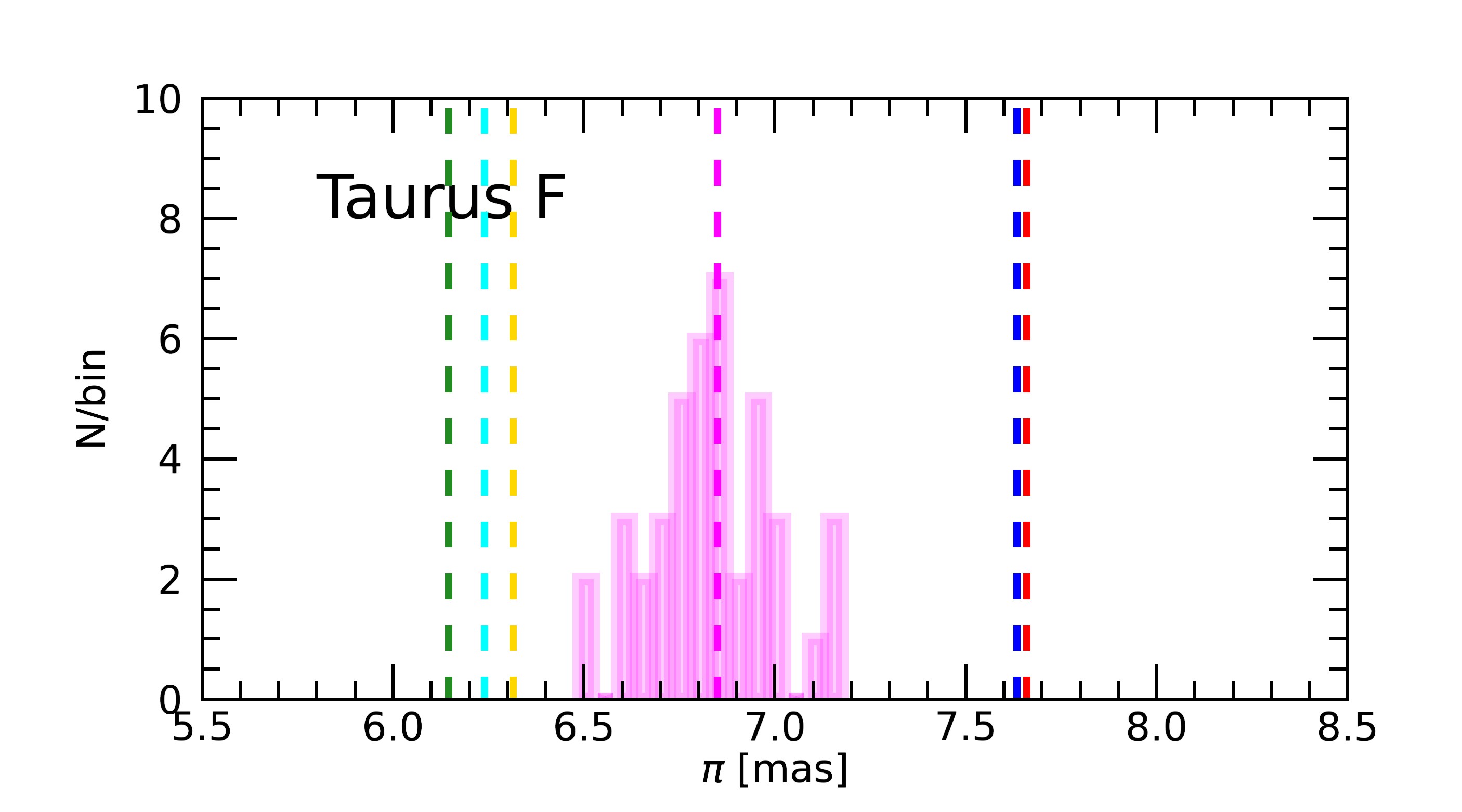}
    \caption{Parallax distribution of the most probable members (P$\ge$80$\%$) of the  Taurus A (red), Taurus B (blue), Taurus C (yellow), Taurus D (green), Taurus E (cyan), and Taurus F (magenta) populations.  
    In each panel the  maximum-likelihood parallaxes for the six populations are also shown as vertical dashed lines using the color code of the histograms.
    }
\label{histplx2}
\end{figure}
\begin{figure}
\centering%
\resizebox{\hsize}{!}{\includegraphics[trim=2cm 6cm 0cm 0cm]{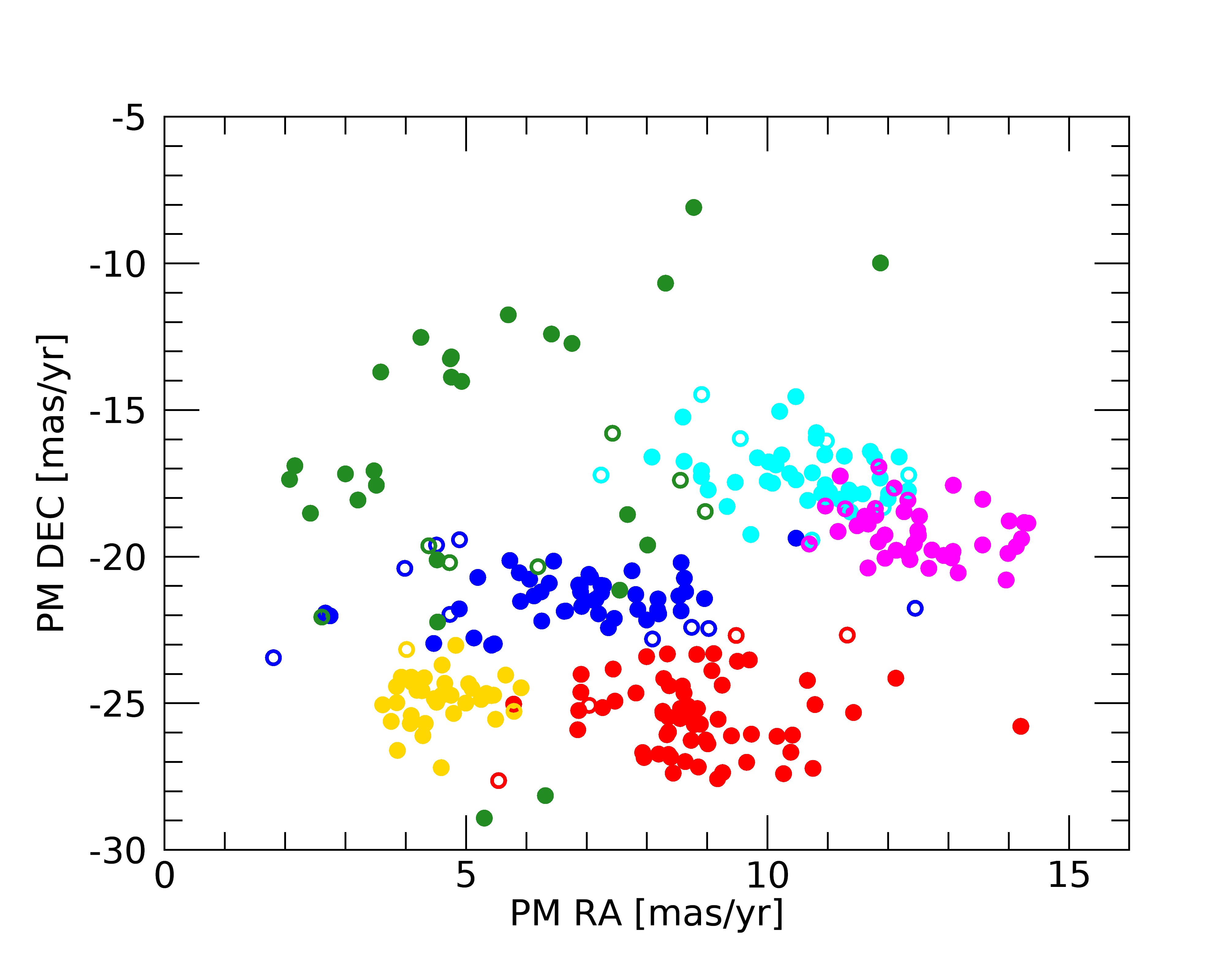}}
\caption{Proper motions in $\alpha$  and $\delta$ of the six populations, color coded as in Fig.~\ref{histplx2}. The filled dots represent sources with probabilities higher than 80\%, while the empty circles represent sources
with probabilities lower than 80\%.
} 
\label{pm2}
\end{figure}

 We fitted our data with a series of models, starting with two populations, and adding one population at a time. We note that each additional population requires that seven new free parameters be fitted (the three centroids of parallax and proper motions and the relative dispersions, plus the fraction of stars).
The statistical significance of each model was tested using the Bayesian information criterion (BIC), which takes into account both the maximum-likelihood and the number of free parameters of the model in order to reduce the risk of overfitting.
The resulting values are shown in Fig.~\ref{BIC}: we see that the BIC decreases up to six populations and then starts to increase again. Therefore, we conclude that the model with six populations is the best fit of our data, given our assumptions.
The best fit parameters for the six populations, 
which we call Taurus A, B, C, D, E and F, are 
listed in Table~\ref{mle-taurus}, while the probability 
of each star  belonging to each population is reported 
in Table~\ref{tb}. Between the six probabilities for each source, the highest identifies the population with which that source is associated. 
We find that 253 of the 283 stars used for the 
calculations have a probability of higher than 80\%  of 
belonging to one of the six populations: 62 to Taurus A, 46 to Taurus B, 36 to Taurus C, 27 to Taurus D, 40 to Taurus E, and 42 to Taurus F. 
The distribution of parallaxes and proper motions of these 253 stars are shown in Figs. \ref{histplx2} and \ref{pm2}, respectively. In Fig.~\ref{herschel} (see Sect.~\ref{sec_herschel}) we show their spatial distribution relative to the molecular cloud traced by {\it Herschel} far-infrared images. \\
We note that the majority of the members of the six populations that have been selected using only proper motions and parallaxes are also well separated in the plane of the sky. 
This is independent confirmation of the results of our analysis.

The relative errors on the derived parallaxes are of the order of 1\% or less, and therefore the bias to the distance calculated by inverting the parallaxes is negligible  \citep[][]{Lurietal2018, Bailer-Jones2015}.
 We find that three of the populations are associated with {\it Taurus Main}: Taurus A and B, located at similar distances (130.5$^{+0.5}_{-0.5}$ pc and 131.0$^{+1.0}_{-1.0}$ pc, respectively), and Taurus E further away at a distance of 160.3$^{+0.6}_{-0.6}$ pc. Taurus C and F are located in the clouds outside {\it Taurus Main}, at 158.4$^{+0.6}_{-0.6}$ pc and  146.0$^{+0.6}_{-0.6}$ pc, respectively. With a mean distance of 162.7$^{+4.4}_{-4.1}$ pc, Taurus D is the least defined population, with a large spread both in astrometric parameters and in the spatial distribution.\\
All the given errors in the parallaxes and distances include only the statistical uncertainties from the fit. At these distances, the conservative systematic error of 0.1 mas discussed by \citet{Lurietal2018} corresponds to $\sim\,$2 pc.

\begin{table*}
\begin{center}
\caption{Results from the maximum-likelihood fit for the parallaxes and proper motions, and relative dispersions, of Taurus A, B, C, D, E and F. 
In the last line we report the distance in parsecs of the six populations computed as described in the text.
\label{mle-taurus}}
\begin{tabular}{l|rrrrrr}
\hline\hline
 \noalign{\smallskip}
& Taurus A & Taurus B & Taurus C & Taurus D & Taurus E & Taurus F\\
  \noalign{\smallskip}
\hline  
\noalign{\smallskip}
$\pi$ [mas]&  7.660 $\pm$ 0.029&  7.634 $\pm$ 0.058&   6.314 $\pm$ 0.025&  6.145 $\pm$ 0.161&  6.240 $\pm$ 0.024&  6.850 $\pm$ 0.029\\
$\mu_\alpha$ [mas/yr]&  8.780 $\pm$ 0.144&      6.883 $\pm$ 0.258&  4.618 $\pm $0.108&  5.572 $\pm$ 0.454&  10.490  $\pm $0.209& 12.413 $\pm$ 0.167       \\
$\mu_\delta$ [mas/yr] & $-$25.373 $\pm$ 0.201& $-$21.380  $\pm$ 0.170& $-$24.842 $\pm$ 0.138& $-$17.368 $\pm$ 0.934& $-$17.172 $\pm$ 0.168& $-$19.077 $\pm$ 0.145   \\
$\sigma_{\pi,0}$ [mas]&  0.192 $\pm$ 0.025& 0.271 $\pm$ 0.058&  0.101 $\pm$ 0.027&  0.663 $\pm$ 0.101& 0.088  $\pm$ 0.034&  0.136 $\pm$ 0.024       \\
$\sigma_{\mu_\alpha,0}$ [mas/yr]&  1.073 $\pm$ 0.113&    1.358 $\pm$ 0.174& 0.591 $\pm$ 0.079& 2.298 $\pm$ 0.314&  1.141 $\pm$ 0.164&  0.976 $\pm$ 0.147\\
 $\sigma_{\mu_{\delta,0}}$ [mas/yr]& 1.268 $\pm$ 0.159&  0.847 $\pm$ 0.123& 0.770 $\pm$ 0.109& 4.738 $\pm$ 0.612&  1.006 $\pm$ 0.126& 0.891 $\pm$ 0.103\\
$f$& 0.233 $\pm$ 0.027& 0.189 $\pm$ 0.028& 0.129 $\pm$ 0.021& 0.127 $\pm$ 0.027& 0.163 $\pm$ 0.024& 0.159 $\pm$ 0.057\\
\noalign{\smallskip}
\hline
\noalign{\smallskip}
d [pc] &130.5$^{+0.5}_{-0.5}$& 131.0$^{+1.0}_{-1.0}$& 158.4$^{+0.6}_{-0.6}$& 162.7$^{+4.4}_{-4.1}$& 160.3$^{+0.6}_{-0.6}$& 146.0$^{+0.6}_{-0.6}$\\
\noalign{\smallskip}
\hline                                
\end{tabular}
\end{center}
\end{table*}


\noindent

\section{Discussion}
\label{discussion}
We organize the discussion as follows. First we compare the distances of the six populations to the literature values. We then investigate the properties and distance of the molecular cloud in order to be able to relate the 3D distribution of the stellar population to the molecular cloud. 
\subsection{Distances of the Taurus stellar populations}
The distance of the Taurus complex reported in the literature \citep[e.g.,][]{PreibischSmith1997} and commonly used so far is consistent with the average parallax of 7.01 mas that we found previously, but this can no longer be considered valid in light of the present results, which show different distances for the six populations, ranging from 130 pc for Taurus A to 160 pc for Taurus C. \\
Several measurements at different wavelengths and using different methods have been taken in recent decades to investigate the distance of the Taurus complex. 
Accurate distances derived from VLBI astrometry are available for several stars.
Values of 128.5$\pm$0.6 pc, 132.8$\pm$0.5 pc, and
132.8$\pm$2.3 pc were found for Hubble 4, HDE 283572,
and the binary V773~Tau, respectively \citep{Torresetal2007,Torresetal2012}, while a distance of 161.2$\pm$0.9 pc was found for the HPTau/G2 system \citep{Torresetal2009}. \citet{Gallietal2018} derived VLBI distances for an additional 18 Taurus members, finding values ranging from 84 to 162 pc. 
We note also
that \citet{Kenyonetal1994} already suggested that the canonical distance of 140$\pm$10 pc was not representative of most stars in this region since they measured a significant depth effect using  optical spectrophotometry of field stars projected on the molecular cloud.
On the contrary, \citet{Dzibetal2018}, using the {\it Gaia} data alone, found that L1495 in Taurus lies at 129 pc,  which corresponds to the distance of Taurus A in our analysis.\\
The presence of multiple populations was also investigated by a recent study of \citet{Luhman2018},
who used colors to identify different populations; we report those colours in Table~\ref{tb}.
We find that all the stars that we classify as part of Taurus A and Taurus B belong to the red population of \citet{Luhman2018}, while stars in Taurus C belong to his cyan population. The blue population corresponds to Taurus E and F, while Taurus D includes sources from all the populations found by \citet{Luhman2018}. 
This is to be 
expected since this is the most spatially dispersed population, without a clear peak in the parallax distribution. We conclude that the populations found by \citet{Luhman2018} are only partially consistent with ours. This is because we adopted  completely different approaches. In his case, starting from the spatial distribution of sources in correspondence to the different clouds, \citet{Luhman2018} characterized their average parallax and proper motions without any additional statistical analysis able to take into account the errors and correlations between astrometric parameters. Recently, \citet{Gallietal2019} 
also studied the Taurus populations,  instead applying a
hierarchical clustering algorithm and removing the outliers using the minimum covariance determinant.
These latter authors found 21 clusters, which are also indicated in Table~\ref{tb}, most of which, however, have less than eight members.
To compare this latter result with the four populations found by \citet{Luhman2018},  \citet{Gallietal2019}  
defined six groups of clusters, which are consistent with the four populations of \citet{Luhman2018}. Comparing our results with the group classification of \citet{Gallietal2019}, we find that Taurus A and B, both red in \citet{Luhman2018}, correspond to Group F in \citet{Gallietal2019},
that Taurus C corresponds to Group C \citep[cyan in][]{Luhman2018}, Taurus E mostly corresponds to Group A, and Taurus F corresponds to Group B  \citep[both blue in ][]{Luhman2018}.

It is important to highlight the fact that our decision to not start our analysis from the spatial distribution of the sources (as is highly recommended in clusters without  a centrally peaked spatial distribution) is motivated by the fact that we wanted to investigate whether or not the molecular cloud, which is mainly structured in filaments, is shaping the spatial distribution of the  different populations.\\

\begin{figure*}
\centering%
\includegraphics[trim=0cm 0cm 0cm 0cm,width=0.9\textwidth]{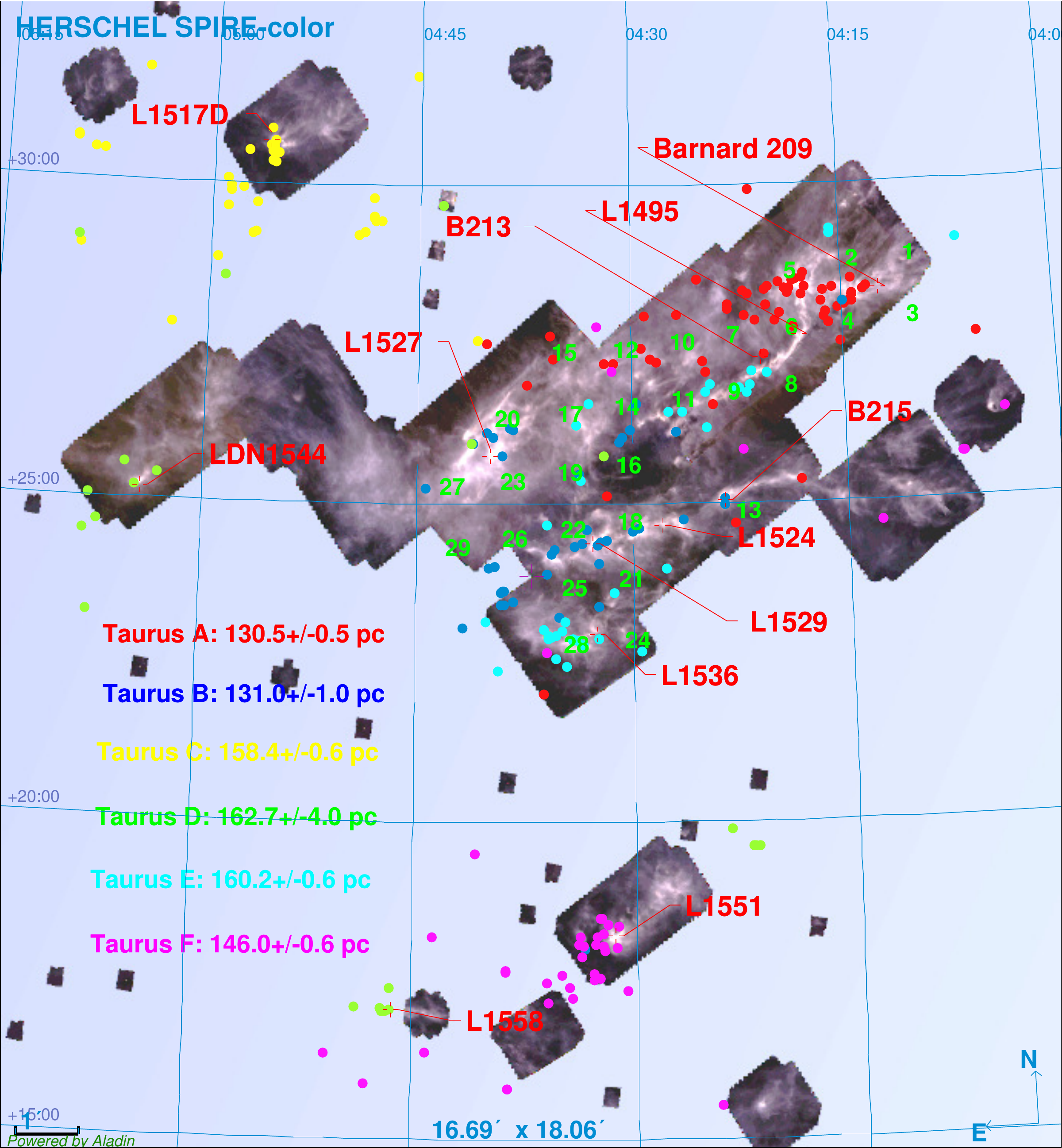}
\caption{{\it Herschel} RGB color composition using all public SPIRE observations from the Herschel Science Archive at 250 $\mu$m (blue), 350 $\mu$m (green), and 500 $\mu$m (red). The filled circles represent the most probable members of Taurus A (in red), Taurus B (in blue), Taurus C (in yellow), Taurus D (in green), Taurus E (in cyan) and Taurus F (in magenta). With {\it Taurus Main} we refer to the Taurus complex excluding the clouds L1517D, LDN 1551 and L1558. The green numbers which appear only on {\it Taurus Main} correspond to the distances in parsec computed in those positions by \citet{Zuckeretal2019}. In red we indicate the names of the clouds in the region. At 150 pc the size of 1$^\circ$ corresponds to 2.6~pc.
}
\label{herschel}
\end{figure*}
\subsection{Taurus molecular clouds and stellar populations}
\label{sec_herschel}
We can now begin to discuss the Taurus star formation history by combining the new {\it Gaia} results with high-resolution observations of the filaments in Taurus from {\it Herschel}. 
A similar approach was used by \citet{Grossscheldetal2018} with  Orion A finding that this is a cometary-like cloud and that the  straight filamentary cloud is only its projection. \\
\noindent
A large-scale continuum map of the Taurus complex was computed from the near-infrared extinction map from \citet{Schmalzletal2010}. 
The {\it Herschel} observations of Taurus were obtained in the context of the Gould Belt survey  \citep{Kirketal2013, Palmeirimetal2013} allowing point-source extraction and the determination of the temperature and density profile of the main filament in the LDN 1495 region. Figure~\ref{herschel} shows a composite {\it Herschel} map of the Taurus complex combining all the public SPIRE observations available 
in the archive \footnote {http://archives.esac.esa.int/hsa/whsa/}. The positions of the Taurus A, B, C, D, E and F populations are also shown, as well as the position of the known clouds in the region.  
We find that, in the {\it Taurus Main} cloud, Taurus A and B are at a similar distance, while Taurus E is located 30~pc further away. The greatest concentration of sources of Taurus A spreads over the B209 
cloud. In the rest of {\it Taurus Main}, all populations follow the filamentary structure traced by the {\it Herschel} maps. Taurus B, in particular, is mostly distributed along the L1524 and L1529 clouds, while Taurus E is projected in the direction of the filamentary clouds L1495 and B213
 and on the cloud L1536.
%
The bulk of the population of Taurus C corresponds to the L1517 cloud, while Taurus F is concentrated on L1551. Few sources of Taurus D are close to the L1544 and L1558 clouds, but in general this population is distributed over the entire complex. A possible explanation for this behavior, given also the spread in parallax and proper motions, is that Taurus D is not a single population but the combination of multiple substructures, which our analysis is not able to distinguish; it may also represent a dispersed young population.
\\

\begin{figure*}[htb]
\centering%
\includegraphics[trim=0cm 0cm 0cm 0cm,width=0.33\textwidth]{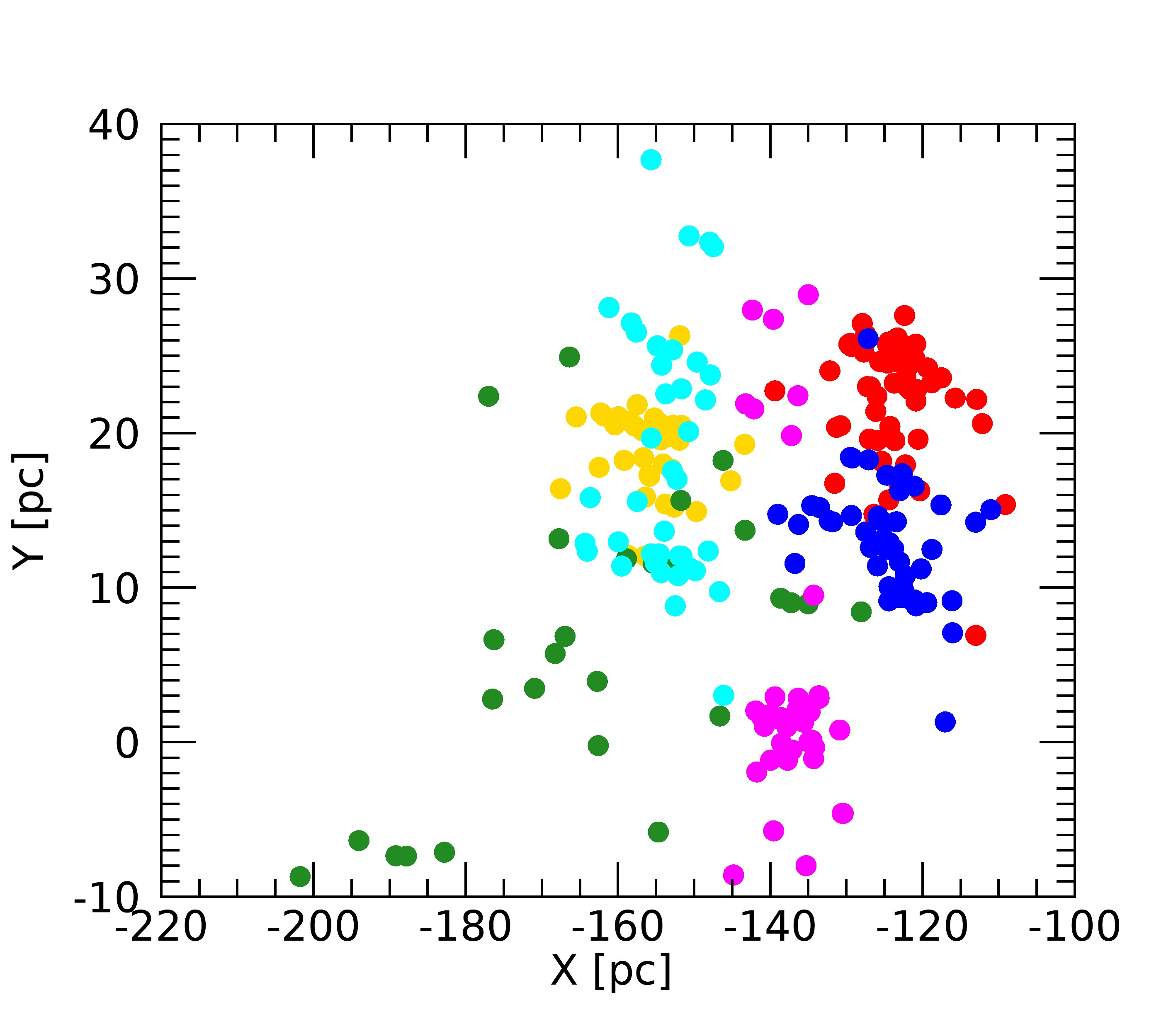}
\includegraphics[trim=0cm 0cm 0cm 0cm,width=0.33\textwidth]{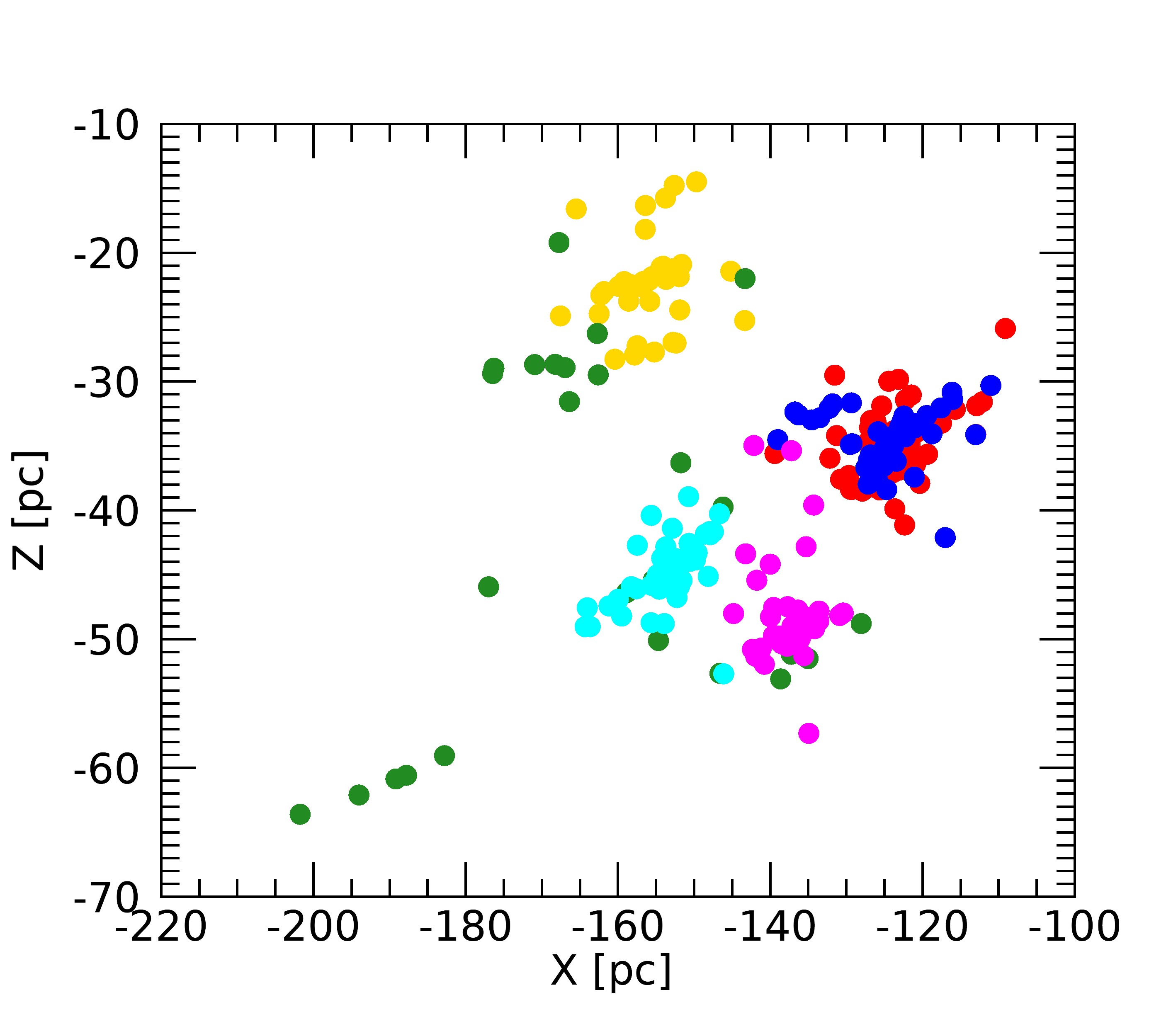}
\includegraphics[trim=0cm 0cm 0cm 0cm,width=0.33\textwidth]{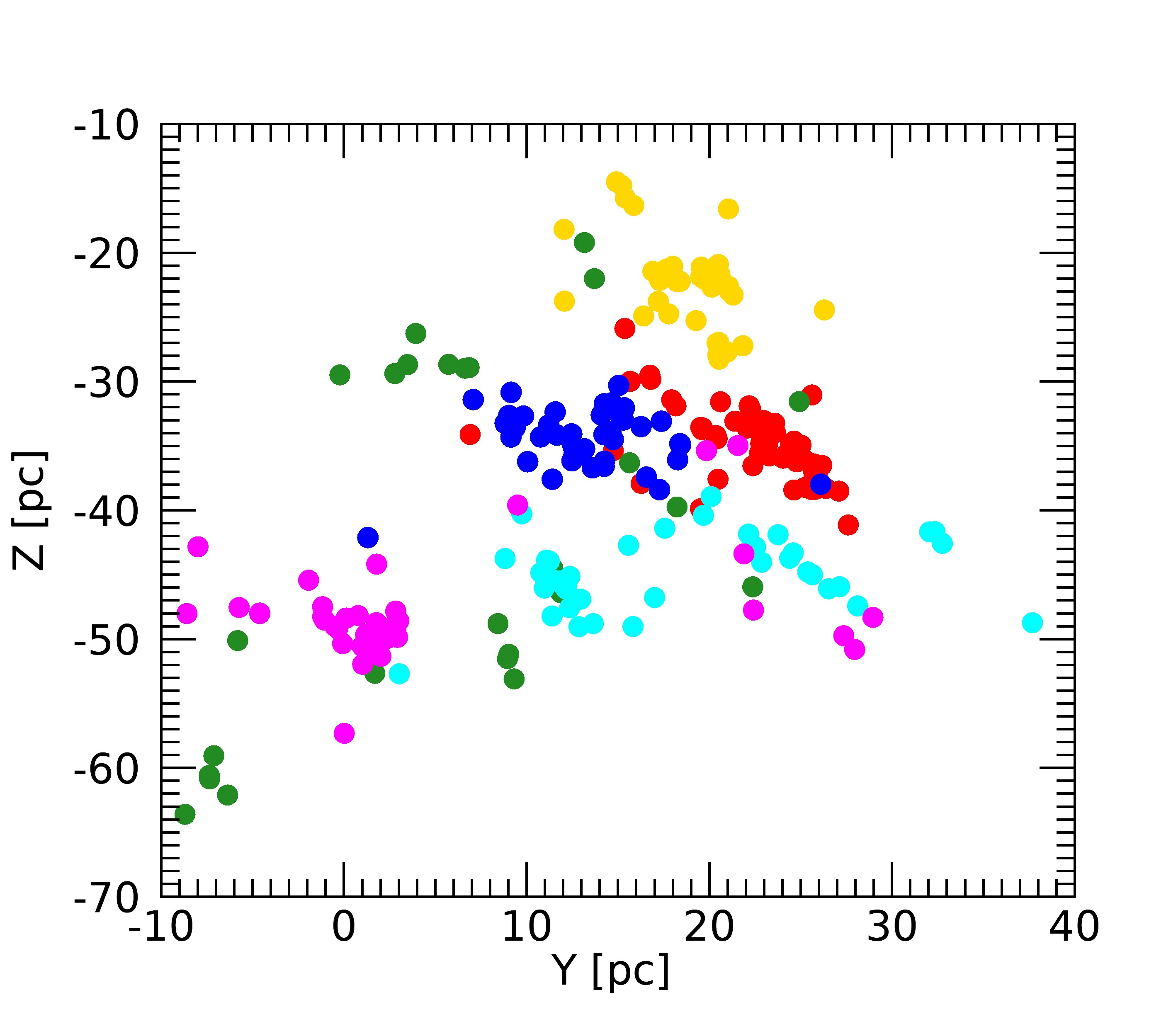}
\caption{Two-dimensional projections of the 3D spatial distribution of the multiple populations of Fig.~\ref{3d} . The color code is as in Fig.~\ref{histplx2}.
}
\label{pos}
\end{figure*}

\noindent
Figure~\ref{3d} shows the 3D spatial distribution of the six 
populations in cartesian galactic coordinates (X,Y,Z), where the X axis 
points towards the galactic center, Y towards the local direction of rotation in the plane of the galaxy, and Z towards the North Galactic Pole. In particular, given 
the galactic coordinates $(l,b)$, 
X,Y,Z are defined as:
\begin{equation}
\begin{array}{l}
X = d\,\cos\,b\,\cos\,l \\
Y = d\,\sin\,l\,\cos\,b \\
Z = d\,\sin\,b.
\end{array}
\end{equation}
The projection of the 3D distribution of the six populations on the individual planes is shown in Fig.~\ref{pos}. We find that Taurus A and B in the (X,Z) projection are compact and overlapped, 
 while in the other two projections they are separated but close; both populations 
extend
for about 20 pc along the X axis and 10 pc along the Y and Z axes. 
In the (X,Y) and (X,Z) planes, Taurus C is 
distributed along three parallel lines separated by a few parsecs and extending for about 10 pc. In the (Y,Z) projection, this population has a more compact structure in the center, which is slightly elongated by about 5~pc in the Y direction. 
Taurus D is also confirmed to be a spread population in 3D. Taurus E has a compact structure of 20$\times$10~pc only in the (X,Z) plane; in the other two projections, it shows 
a compact structure of 20$\times$5~pc and 5$\times$10~pc in the (X,Y) and (Y,Z) planes, respectively, which corresponds to the L1536 cloud. This structure is connected by a few sources along 10~pc to a 
 linear 
distribution of sources of 10~pc in length corresponding to the L1495 filament. 
The bulk of sources of Taurus F 
are concentrated within 10~pc in X and 5~pc in Y and Z; this corresponds to the L1551 cloud. A few additional sources are 
scattered over 20~pc in X and Z, and  $\sim$40~pc in Y.   
\\
\noindent
The distance of the molecular cloud can 
be estimated using the field stars in and outside of the cloud from {\it Gaia} DR2. 
This was done recently by \citet{Yanetal2019}, who found a single distance of 145$^{+12}_{-16}$ pc for the entire molecular cloud. 
A similar approach was adopted by
\citet{Zuckeretal2019}, who compiled  a uniform catalog of distances to local molecular clouds. The methodology of these latter authors, based on the work of \citet{Schlaflyetal2014}, infers
a joint probability distribution function on distance and reddening for individual stars based on optical and near-infrared  photometric surveys and {\it Gaia} parallaxes, and also on modeling of the cloud as a simple dust
screen to bracket the dust
screen between unreddened foreground stars and reddened background stars. 
 The work of 
\citet{Zuckeretal2019} allowed the authors to not only find  a single distance for the surface of the Taurus molecular cloud of 141$\pm$9 pc,
  but also to map the 3D surface of the region previously defined as {\it Taurus Main}. 
 \begin{table}
\begin{center}
\caption{Local distances in {\it Taurus Main} computed by
\citet{Zuckeretal2019} with: the number appearing in Fig.~\ref{herschel} (col.1 \& 4),
galactic positions (col.2 \& 5), and the distances  (col.3 \& 6). \label{z19} }
\begin{tabular}{lcl | lcl}
\hline
\noalign{\smallskip}
\#&     l [$^\circ$] , b [$^\circ$]     &       D [pc] &\#&     l [$^\circ$] , b [$^\circ$]  &       D [pc]\\
  \noalign{\smallskip}
\hline  
\noalign{\smallskip}
1       &       167.3,  -16.3   &       133$^{+3}_{-2   }$&     16      &       173.0,    -15.1   &       139$^{+16}_{-14 }$\\
2       &       168.0,  -15.7   &       116$^{+17}_{-15 }$&     17      &       173.0,    -13.9   &       127$^{+5}_{-4   }$\\
3       &       168.0,  -17.0   &       132$^{+6}_{-7   }$&     18      &       173.7,    -15.7   &       142$^{+8}_{-4   }$\\
4       &       168.8,  -16.3   &       136$^{+8}_{-5   }$&     19      &       173.7,    -14.5   &       130$^{+21}_{-4  }$\\
5       &       168.8,  -15.1   &       136$^{+13}_{-10 }$&     20      &       173.7,    -13.2   &       123$^{+11}_{-6  }$\\
6       &       169.5,  -15.7   &       132$^{+2}_{-2   }$&     21      &       174.4,    -16.3   &       150$^{+7}_{-6   }$\\
7       &       170.2,  -15.1   &       142$^{+3}_{-4   }$&     22      &       174.4,    -15.1   &       160$^{+6}_{-8   }$\\
8       &       170.2,  -16.3   &       162$^{+10}_{-9  }$&     23      &       174.4,    -13.9   &       137$^{+6}_{-4   }$\\
9       &       170.9,  -15.7   &       152$^{+4}_{-12  }$&     24      &       175.1,    -17.0   &       150$^{+7}_{-6   }$\\
10      &       170.9,  -14.5   &       114$^{+10}_{-6  }$&     25      &       175.1,    -15.7   &       152$^{+5}_{-6   }$\\
11      &       171.6,  -15.1   &       128$^{+3}_{-2   }$&     26      &       175.1,    -14.5   &       161$^{+10}_{-9 }$\\
12      &       171.6,  -13.9   &       154$^{+4}_{-9   }$&     27      &       175.1,    -13.2   &       150$^{+7}_{-6   }$\\
13      &       172.3,  -17.0   &       151$^{+4}_{-2   }$&     28      &       175.8,    -16.3   &       159$^{+2}_{-1   }$\\
14      &       172.3,  -14.5   &       147$^{+4}_{-5   }$&     29      &       175.8,    -13.9   &       162$^{+5}_{-4   }$\\
15      &       172.3,  -13.2   &       126$^{+6}_{-5   }$&     & & \\
\noalign{\smallskip}
\hline                        
\hline                        
\end{tabular}
\end{center}
 \end{table}
This discussion can be taken a step further by 
  considering the 29 local distances computed by \citet{Zuckeretal2019} over {\it Taurus Main}  at the positions indicated with numbers 
 1--29 in the central part of  Fig.~\ref{herschel}. The values of the corresponding distances are given in Table~\ref{z19}.
  In the following, we take into account the measurement errors on both the stellar populations and the molecular cloud when comparing their relative positions. Most of the sources of Taurus A lie between the B209 and L1495 
 clouds: the stellar population, at 130 pc, is associated with the molecular cloud, which ranges between 132 and 136 pc. Taurus B extends over a filamentary structure, identified by a first filament corresponding to B215 and a second filament corresponding to L1524 and L1529. In this case, the stellar population is located in front of both filaments at a range of distances  between 4 and 20 pc.  A further six sources of Taurus B correspond instead to the L1527 cloud, and are associated with it, within the errors. \\
The part of Taurus E which in 2D appears to be associated with the L1495 
 and B213 filamentary structures is instead more distant by 2$-$20 pc at the center of the filamentary structure, and by 25$-$35 pc at its ends.
The rest of Taurus E is on the surface of the L1536 cloud, at the same distance of about 160 pc. \\
Outside of {\it Taurus Main}, the distances of the  individual clouds are not known. 
However, 
we find that Taurus C, at a distance of $\sim$160 pc, is spatially distributed on the L1517 cloud,
 while Taurus F is spatially distributed on L1551 and located at a distance of $\sim$146 pc, consistent with the mean distance of the Taurus molecular cloud. 
Assuming we have the same situation as in the case of L1527 and L1536, where the distances of the stellar populations correspond to the measured distances of the clouds, this might suggest that L1517 and L1551 lie at $\sim$160 pc and  $\sim$146 pc, respectively, as do their associated stellar populations.
\\
In this scenario, part of the different populations are still associated to the molecular cloud and the dynamical interactions between the stellar populations and the molecular cloud are still ongoing. 
It is important to note that the distances found for the molecular cloud represent only its upper layer closer to us. 
 While the stellar population seems to be always associated with the clouds, this is not the case when filamentary structures are present.  \\
A possible explanation is that the stellar population is moving away from the filamentary molecular cloud. Another possibility is that a physical mechanism  actively removed the molecular cloud from the cluster. 
A further dedicated study is required to decipher which of the above mentioned scenarios has taken place in the Taurus complex, and to investigate the relation between star and clump formation and filaments in this region.

\subsection{Kinematics of the Taurus Main populations}
In the above discussion we conclude that only part of the stellar populations and the molecular cloud are still associated. 
 In this section, we concentrate on the kinematics of the three populations associated with {\it Taurus Main}, namely Taurus A, B, and E. 
\begin{figure*} 
\centering%
\includegraphics[trim=0cm 0cm 0cm 0cm,width=0.33\textwidth]{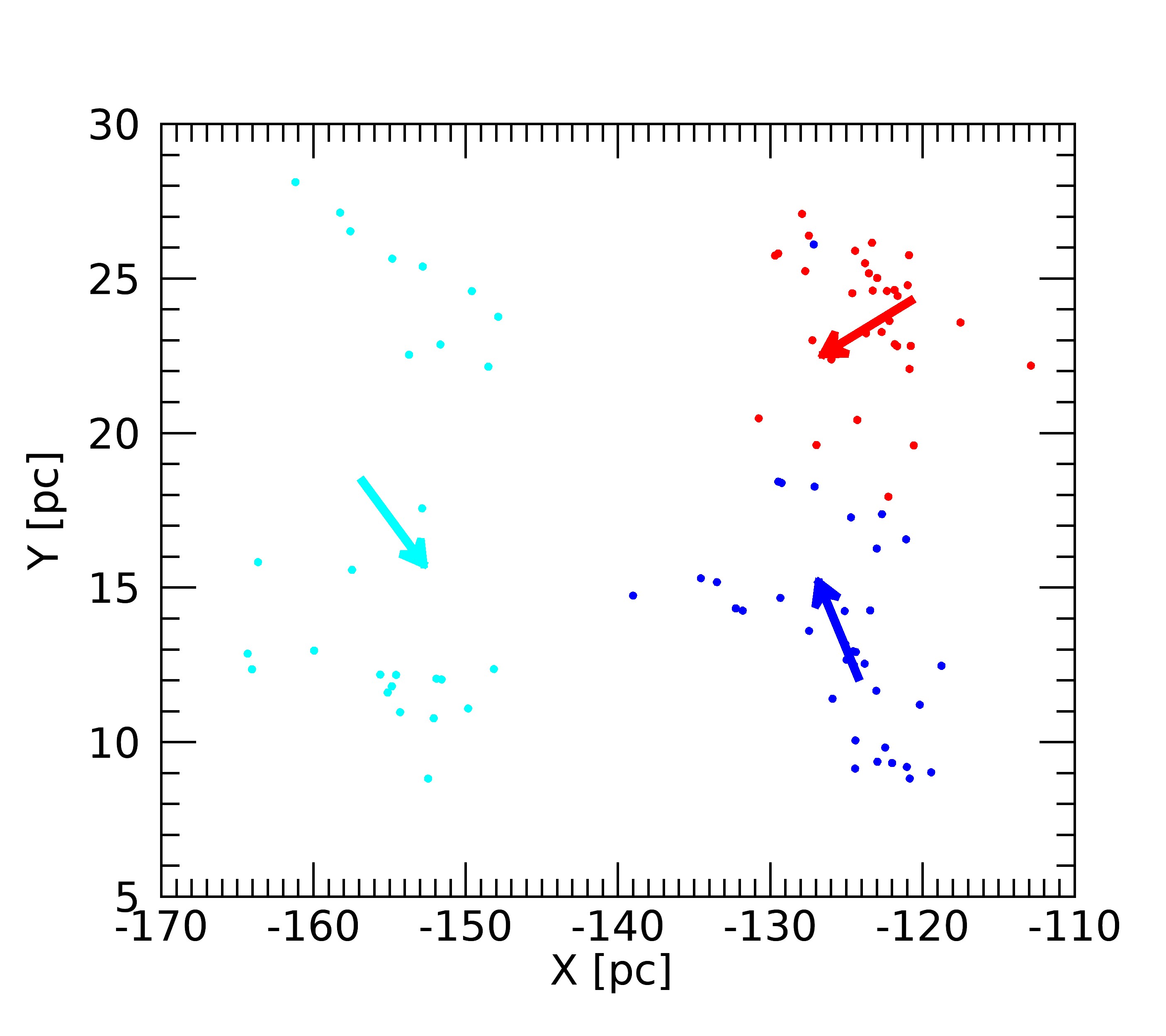}
\includegraphics[trim=0cm 0cm 0cm 0cm,width=0.33\textwidth]{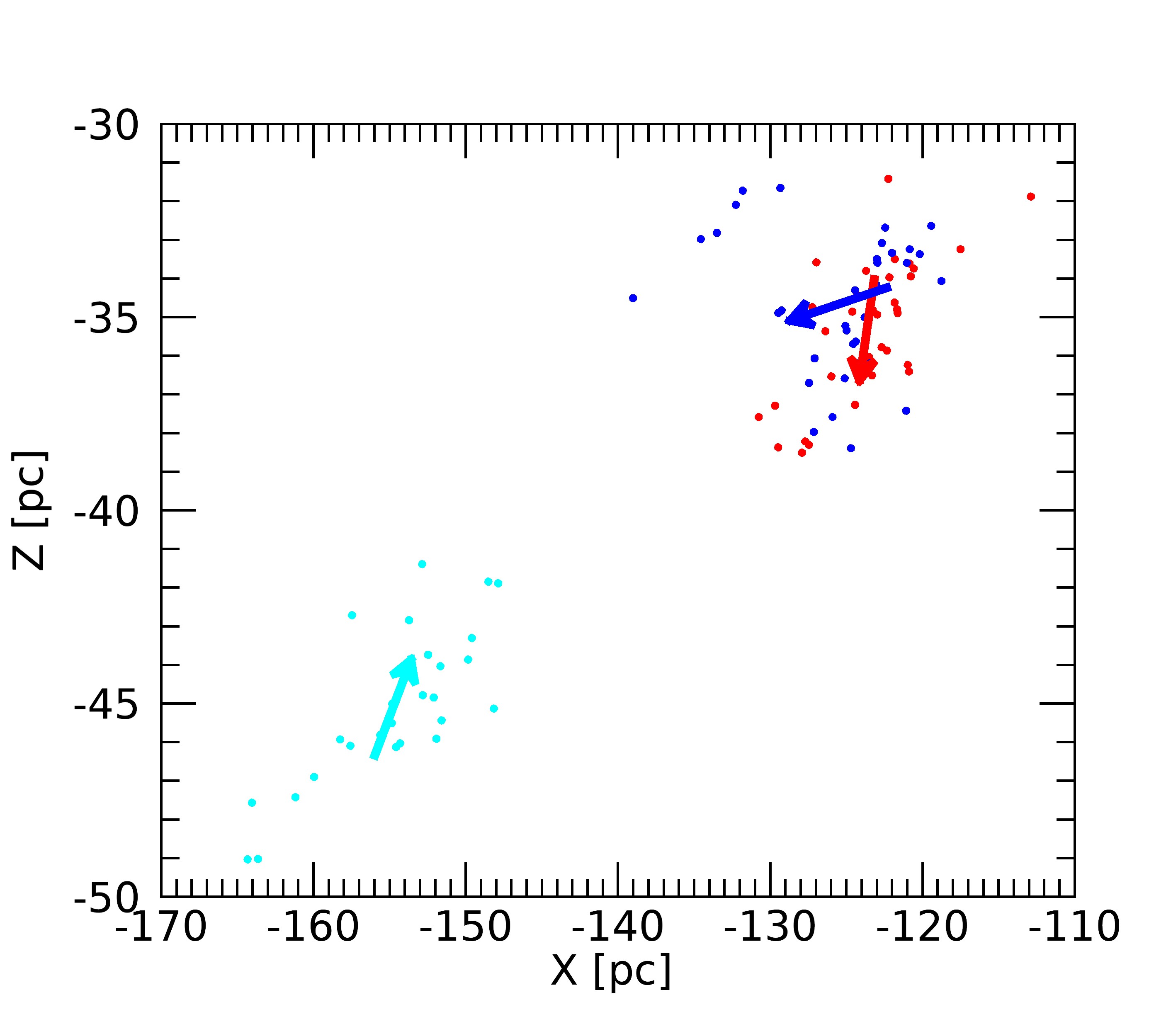}
\includegraphics[trim=0cm 0cm 0cm 0cm,width=0.33\textwidth]{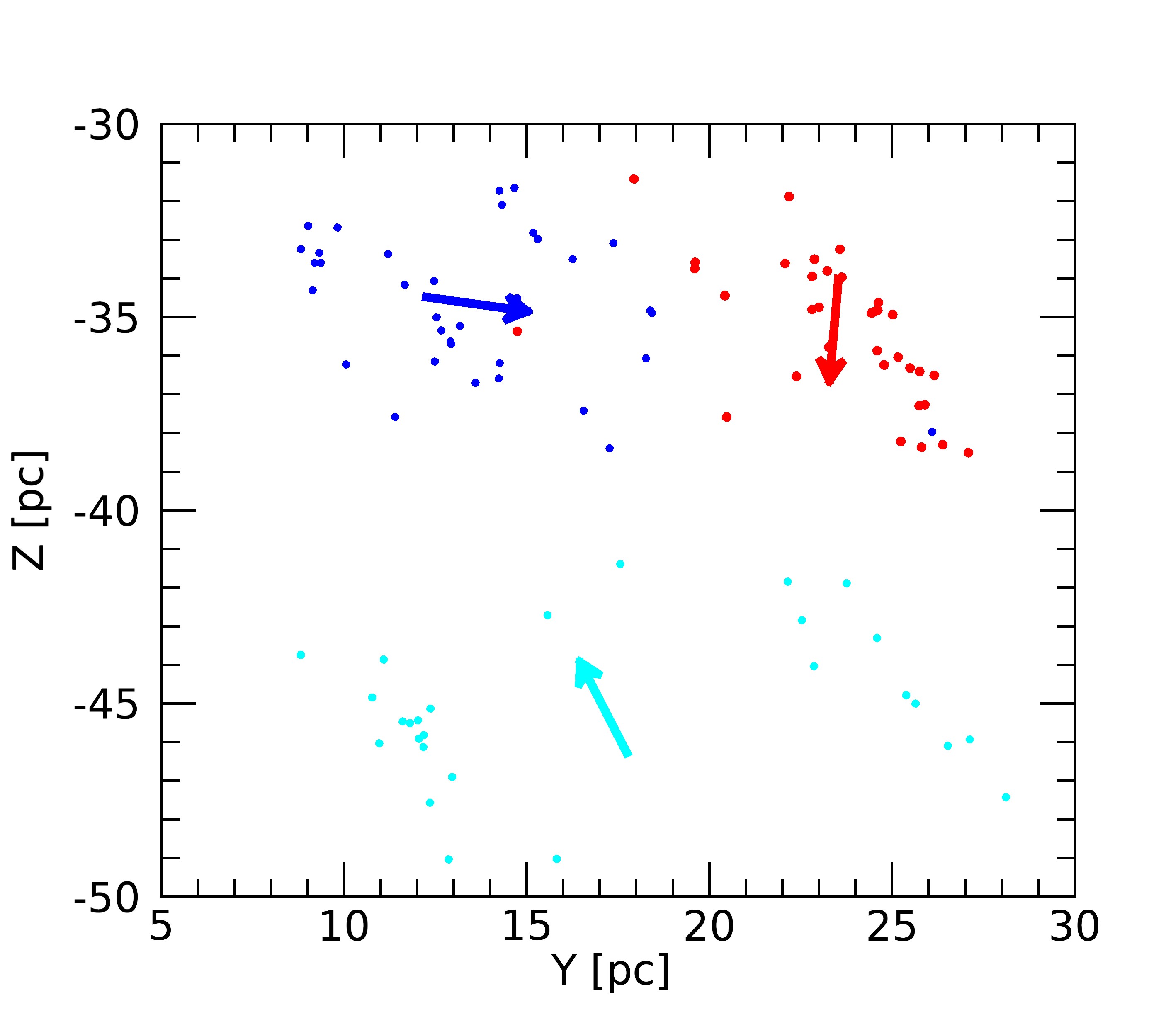}
    \caption{ Spatial distribution (as in Fig.~\ref{pos}) of Taurus A (red), B (blue), and E (cyan),  represented by the small dots, with RV measurements. The 
    arrows represent the differential space velocities (U,V,W) with respect to the mean motion between all the sources of the three populations. Each differential value has been magnified by 100.}
\label{uvw}
\end{figure*}
 We use the available radial velocities
collected from the literature  by \citet{Gallietal2019}. 
All three populations have radial velocity measurements for more than 20 stars; the corresponding histograms are shown in Fig.~\ref{histrv}. 
Although the accuracy of these measurements does not allow a proper investigation of possible subclusterings, we can in any case see that, while Taurus A and B share a similar peak of the distribution, the peak of Taurus E is shifted by $\sim$1 km/s.\\ 
We computed the galactic space velocity (U, V, W) 
\footnote{ We used the implementation of the gal\_uvw.pro IDL routine available in the public {\it astrolib} library, presented by \citet{Gagneetal2014}}
in the same reference system as the (X,Y,Z) positions computed in Sect.~\ref{sec_herschel}. 
These velocities have also been corrected for the solar motion 
\footnote{For the solar motion we adopt the value of (U,V,W)$\odot$ = (-8.5, 13.38, 6.49) from \citet{Coskunogluetal2011}.} to the local standard of rest.
We considered the mean 3D velocities between Taurus A, B, and E 
as a reference to 
compute the mean differential velocities of each population.  
The results are shown in 
Fig.~\ref{uvw} with the three projections of the 3D average motions on the planes (X,Y), (X,Z), and (Y,Z). 
As discussed above, Taurus A and B are overlayed in the (X,Z) projection, and the relative motions can be misleading. In the (X,Y) projection, Taurus A and B are converging to a common point, while in the (Y,Z) projection, Taurus B is moving in the direction of Taurus A. This might be evidence that the two populations are merging. In two out of three projections, Taurus E appears to be approaching Taurus A and B. \\
Additional high-precision radial velocities of each member, in combination with astrometric data, are required to understand the dynamics of the entire region. 
\noindent

\section{Summary and Conclusions}
\label{concl}

We carried out a statistical analysis (taking into account errors and covariances between the parameters) of the Taurus members with a reliable {\it Gaia} DR2  counterpart in order to look for multiple populations in the Taurus complex. 
Our results allow us to infer detailed information on the relation between the stellar population and the molecular cloud structures in the {\it Herschel} maps.\\
We find six stellar populations.
Three of them are located over {\it Taurus Main}: Taurus~A and B  at a similar distance of $\sim$ 130 pc, and Taurus~E at $\sim$ 160 pc. 
We find that these stellar populations lie at a similar distance to the molecular cloud when this is structured as a cloud, suggesting that they are associated. This is not the case when the molecular cloud is filamentary.
 Taurus C and F are mostly concentrated on the L1517 and L1551 clouds.
The sixth population, Taurus D, is widely spread spatially and kinematically, suggesting it could either be a sparse young population, or be composed of multiple substructures.\\
 The analysis of the differential average velocity 
 suggests that Taurus A and B might 
be 
merging, while Taurus E is proceeding towards Taurus A and B.
However, a definitive conclusion on their kinematics will only be possible when accurate radial velocities become available for all the members.

This study supports the view that star formation occurred in clumpy and filamentary structures that are evolving independently, and that Taurus is not the result of the expansion
of a single star-formation episode. 

\appendix
\section{Additional material}
\begin{figure*}[htb]
\centering%
\resizebox{\hsize}{!}{\includegraphics[clip,trim=10cm 5cm 10cm 30cm, width=0.7\hsize] {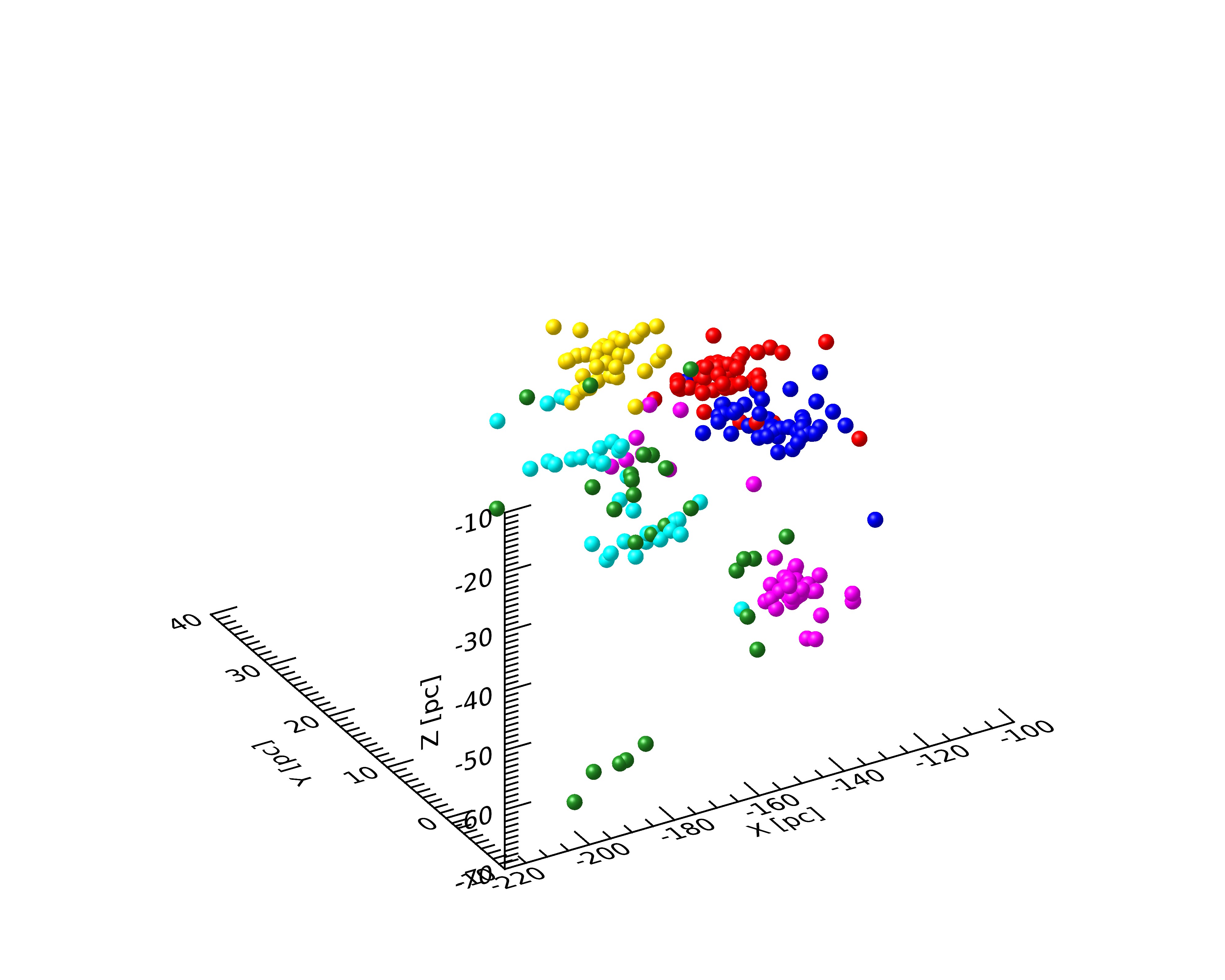}}
\caption{Three-dimensional spatial distribution of the multiple populations centered around the  galactic center. The color code is as in Fig.~\ref{histplx2}. An animation with different orientation of this plot is available in the online version of the paper. }
\label{3d}
\end{figure*}

\begin{figure*}
\centering%
\includegraphics[trim=0cm 0cm 1cm 1cm,width=0.47\textwidth]{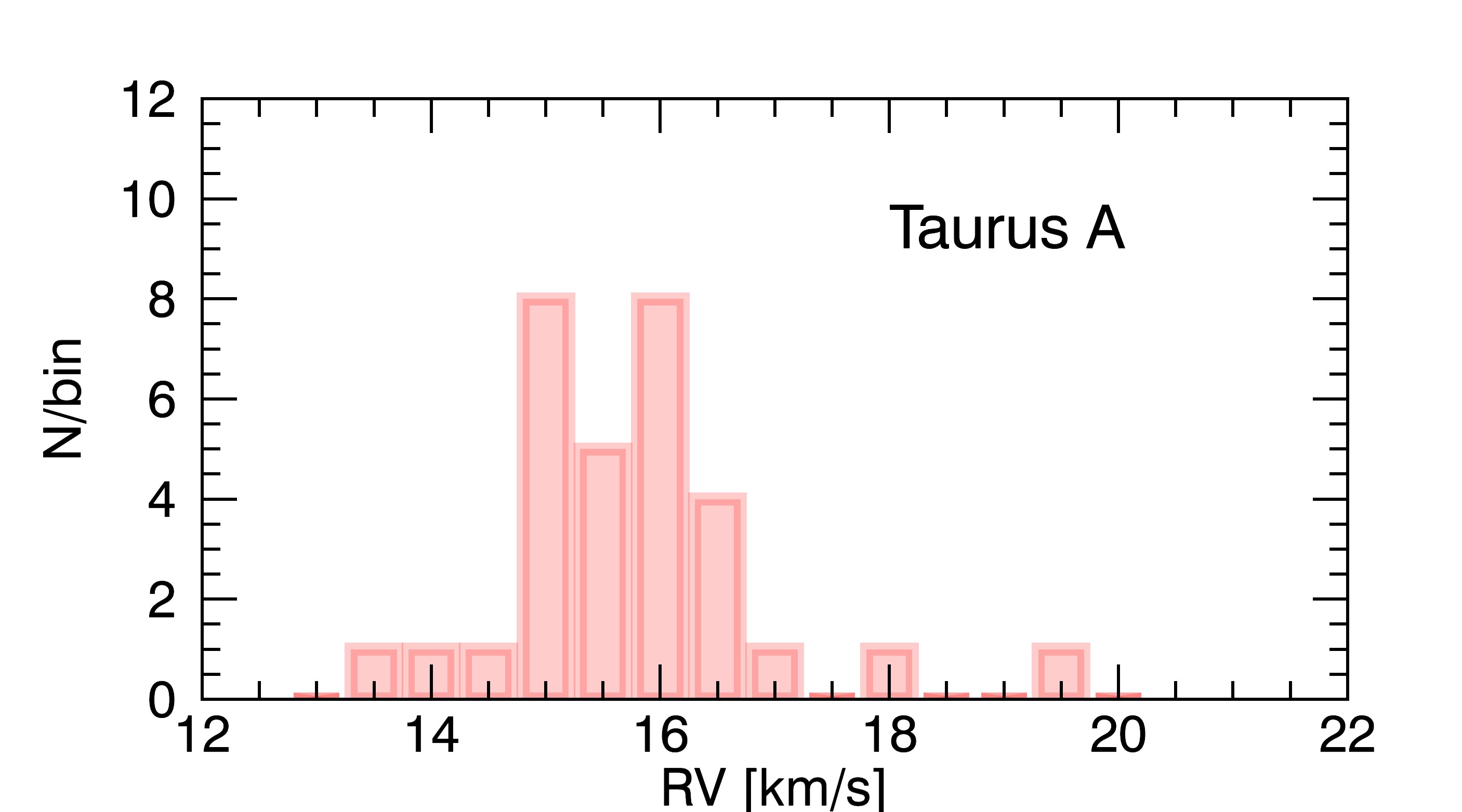}
\includegraphics[trim=0cm 0cm 1cm 1cm,width=0.47\textwidth]{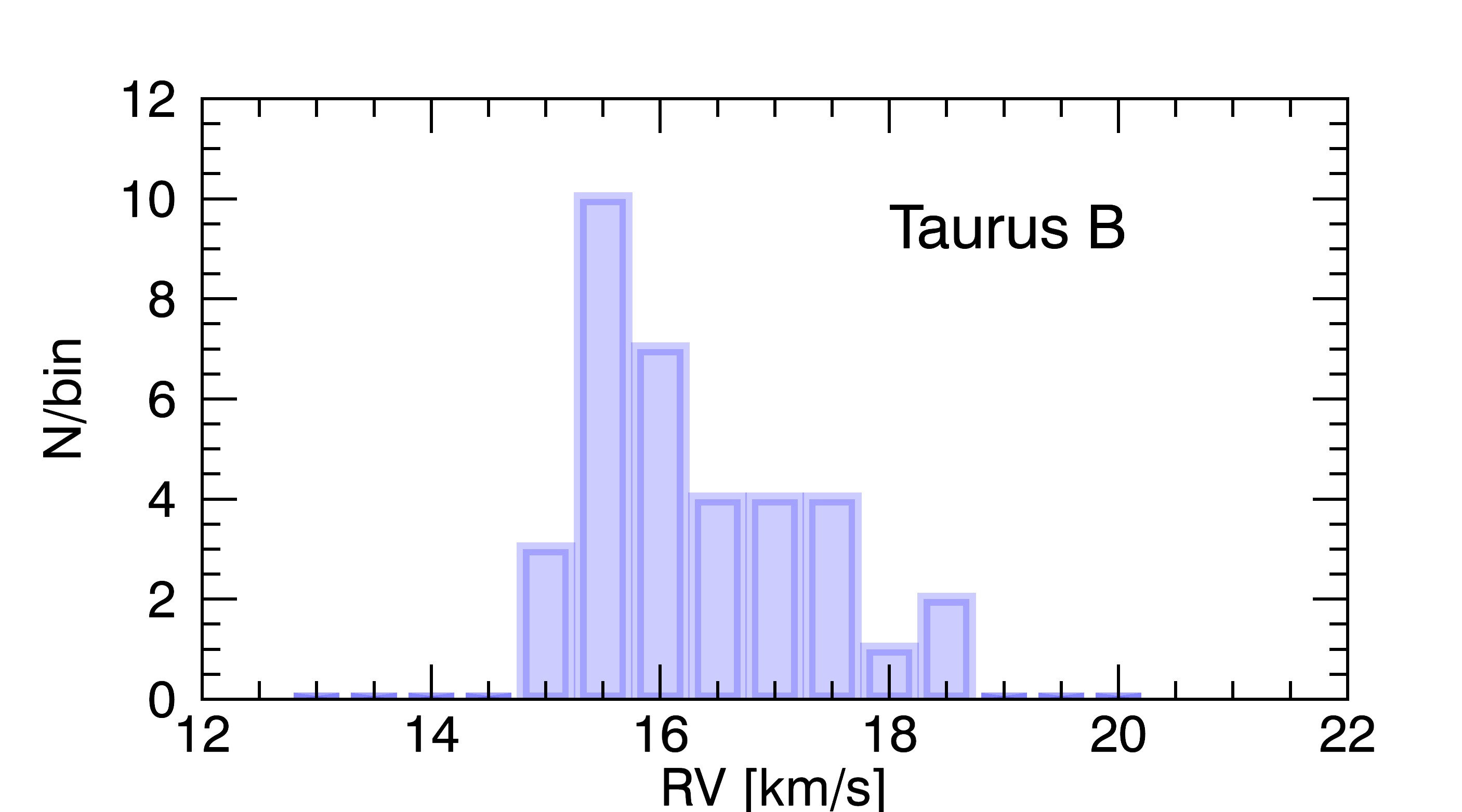}

\includegraphics[trim=0cm 0cm 0cm 0cm,width=0.47\textwidth]{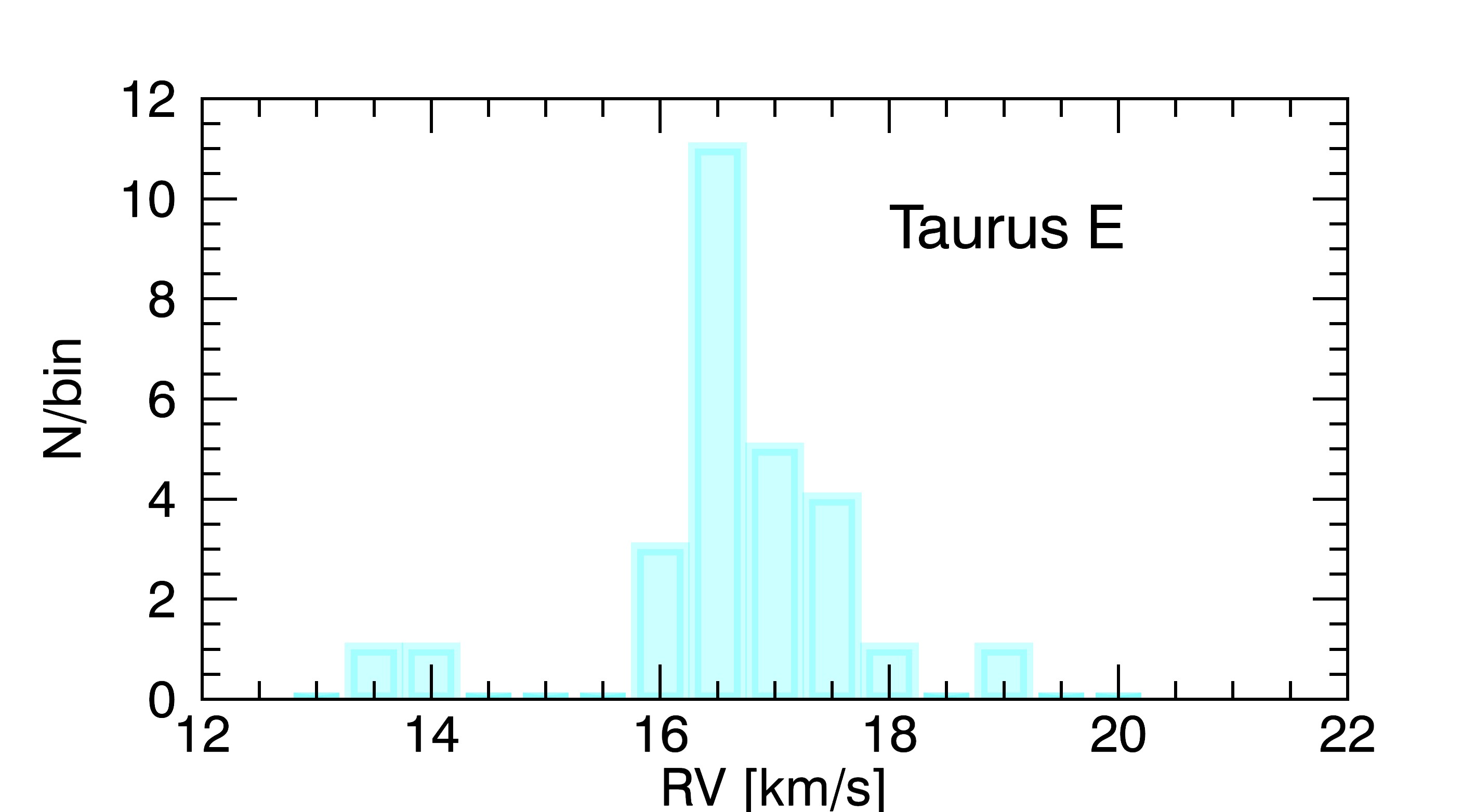}
    \caption{ Distribution of radial velocities from \citet{Gallietal2019} for the most probable members of the  Taurus A (red), Taurus B (blue), and the Taurus E (cyan) populations.
    }
\label{histrv}
\end{figure*}

 Here, Table~\ref{tb}  provides a reduced version of the complete table of membership probabilities; the complete table is available online. Figure~\ref{3d} shows the 3D spatial distribution of the six populations, which  can be rotated in the electronic version of the paper. The histograms  in Fig.~\ref{histrv} show the distribution of the radial velocity compiled from the literature for Taurus A, B, and E.
\begin{sidewaystable*}
\caption{Reduced first ten lines of the table available in the online
version. RA and DEC positions and the astrometric data are from {\it Gaia},
the spectral types and the colors in the ``pop'' column are from
\citet{Luhman2018}, the ``clust'' and ``mem'' columns are the cluster 
number and membership from \citet{Gallietal2019}. \label{tb}}
\centering
\small
\begin{tabular}{lrccccccccc}
\hline\hline
\noalign{\smallskip}
Name     & Gaia DR2 ID&  RA         & DEC        & PLX     & E\_PLX  &  PMRA    & E\_PMRA & PMDE      & E\_PMDE&  \\
\noalign{\smallskip}
\hline
\noalign{\smallskip}
AB Aur   &  156917493449670656& 73.9410446 & 30.5510888 & 6.13996 & 0.05709 &  3.92615 & 0.09673 & -24.11163 & 0.06753 & \ldots\\
CX Tau   &  162758236656524416& 63.6994652 & 26.8029627 & 7.81673 & 0.03963 &  9.02460 & 0.10009 & -22.45099 & 0.06506 & \ldots\\
DL Tau   &  148010281032823552& 68.4128640 & 25.3438374 & 6.27593 & 0.04767 &  9.32913 & 0.08686 & -18.29044 & 0.06509 & \ldots\\
FM Tau   &  163184366130809472& 63.5566418 & 28.2135524 & 7.57948 & 0.04656 &  8.58744 & 0.10322 & -24.41631 & 0.06664 & \ldots\\
FT Tau   &  149623711269425408& 65.9133224 & 24.9371978 & 7.82449 & 0.05194 &  6.91801 & 0.11914 & -21.69847 & 0.07524 & \ldots\\
GM Tau   &  148449845165337600& 69.5889388 & 26.1537301 & 7.22990 & 0.14906 &  5.46909 & 0.26974 & -22.97363 & 0.21549 & \ldots\\
GO Tau   &  148106316500918272& 70.7628400 & 25.3384431 & 6.91711 & 0.04803 &  4.72669 & 0.10092 & -20.20513 & 0.04601 & \ldots\\
HD 28354 &  152189662169148288& 67.3326699 & 27.4041113 & 7.28903 & 0.10899 &  7.26521 & 0.17405 & -25.14882 & 0.13331 & \ldots\\
LkCa 15  &  144936836795636864& 69.8241791 & 22.3508662 & 6.29465 & 0.04814 & 10.47118 & 0.12711 & -17.38300 & 0.05969 & \ldots\\
LkCa 19  &  156900622818205312& 73.9040623 & 30.2985351 & 6.26297 & 0.07864 &  4.30736 & 0.15384 & -24.13216 & 0.07560 & \ldots\\
\noalign{\smallskip}
\hline
\end{tabular}
\begin{tabular}{ccccccccrllrc}
\noalign{\smallskip}
\hline\hline
     & ruwe& $p_A$ & $p_B$ & $p_C$ & $p_D$ & $p_E$ & $p_F$ & RV& SpType & pop  & clust& mem\\
\noalign{\smallskip}
\hline
\noalign{\smallskip}
\ldots   & 1.02& 0.00& 0.00& 0.98& 0.02& 0.00& 0.00&   8.90& A0     & cyan &  1& 1 \\
\ldots   & 1.09& 0.41& 0.59& 0.00& 0.00& 0.00& 0.00&  16.63& M2.5   & red  &  7& 0 \\
\ldots   & 1.16& 0.00& 0.00& 0.00& 0.01& 0.99& 0.00&  13.94& K5.5   & blue & 13& 1 \\
\ldots   & 1.27& 1.00& 0.00& 0.00& 0.00& 0.00& 0.00& \ldots& M4.5   & red  &  7& 1 \\
\ldots   & 1.09& 0.00& 0.99& 0.00& 0.00& 0.00& 0.00&  17.24& M3     & red  &  9& 1 \\
\ldots   & 1.23& 0.01& 0.91& 0.00& 0.08& 0.00& 0.00&  16.46& M5     & red  & 14& 0 \\
\ldots   & 1.09& 0.00& 0.24& 0.00& 0.76& 0.00& 0.00&  15.42& M2.3   & red  & 15& 1 \\
\ldots   & 1.10& 0.99& 0.00& 0.00& 0.01& 0.00& 0.00& \ldots& B9     & red  &  7& 0 \\
\ldots   & 1.18& 0.00& 0.00& 0.00& 0.00& 1.00& 0.00&  17.65& K5.5   & blue & 18& 0 \\
\ldots   & 1.02& 0.00& 0.00& 1.00& 0.00& 0.00& 0.00&  13.58& K2     & cyan &  1& 1 \\
\noalign{\smallskip}
\hline
\noalign{\smallskip}
\end{tabular}
\end{sidewaystable*}

\begin{acknowledgements}
We thank the referee for his/her comments who helped to significantly improve the paper. This project has received funding from the European Union's Horizon 2020 research and innovation programme under the Marie Sklodowska-Curie 
grant agreement No 664931. 
This work has made use of data from the European Space Agency (ESA)
mission {\it Gaia} (\url{https://www.cosmos.esa.int/gaia}), processed by
the {\it Gaia} Data Processing and Analysis Consortium (DPAC,
\url{https://www.cosmos.esa.int/web/gaia/dpac/consortium}). Funding
for the DPAC has been provided by national institutions, in particular
the institutions participating in the {\it Gaia} Multilateral Agreement. VR acknowledges the inspiring discussions with Steve. 
\end{acknowledgements}
\bibliographystyle{aa}
\bibliography{references}
\end{document}